\documentclass[11pt, oneside, reqno]{amsart}   	
\usepackage{geometry}                		
\usepackage{amssymb}
\usepackage{mathrsfs}
\begin{document}
\begin{center}
Commutative deformations of general relativity: nonlocality,
causality, and dark matter
\\ 
\vspace{2ex}
P. G. N. de Vegvar \\ 
\vspace{2ex}
SWK Research\\
1438 Chuckanut Crest Dr., Bellingham, WA 98229, USA\\
  \vspace{2ex}
Paul.deVegvar@post.harvard.edu\\ 
\vspace{2ex}
\end{center}
Hopf algebra methods are applied to study Drinfeld twists of (3+1)-diffeomorphisms and deformed general relativity on \emph{commutative} manifolds.
A classical nonlocality length scale is produced above which microcausality emerges.
Matter fields are utilized to generate self-consistent Abelian Drinfeld twists in a background independent manner 
and their continuous and discrete symmetries are examined.
There is negligible experimental effect on the standard model of particles. 
While baryonic twist producing matter would begin to behave acausally for rest masses above $\sim1-10$ TeV,
other possibilities are viable dark matter candidates or  a right handed neutrino. 
First order deformed Maxwell equations are derived and yield immeasurably small cosmological dispersion 
and produce a propagation horizon only for
photons at or above Planck energies. This model incorporates dark matter without any appeal to extra dimensions, supersymmetry,
strings, grand unified theories, mirror worlds, or modifications of Newtonian dynamics.
\vspace{2.8in}
\pagebreak

\section{Introduction and Motivation}
For several decades noncommutative manifolds have been an active area of mathematics research and many of these methods have been adopted by theoretical
physicists to study the quantum properties of spacetime at the Planck scale \cite{Ahlu}\cite{Dopplicher}\cite{Douglas}\cite{Chamseddine}.
Ideas from Hopf algebras, deformed diffeomorphisms, and quantum Lie algebras
have been utilized to investigate models of quantum spacetime and field theory where the coordinates $x^{\mu }$ are promoted to noncommuting operators 
obeying $[\hat{x}^{\mu }, \hat{x}^{\nu }] = i \hat\theta ^{\mu\nu}$. There have also been studies of non-commutatively 
deformed classical spacetimes \cite{Aschieri_2}\cite{Schenkel},
which introduce a nonlocal star product of objects living on a still classical manifold, meaning $f\star g \ne g \star f$.
Here we apply Hopf algbras to commutatively deformed curved classical manifolds, where  $f\star g = g \star f$.
The commutative deformation approach to classical manifolds has just recently begun 
to be explored by  researchers \cite{Lizzi}\cite{Galluccio}\cite{Ardalan} who studied flat spacetime. 
In this work the Hopf algebra approach is physically motivated by recent studies \cite{PdV} about 
how \emph{background independent} theories of canonical quantum gravity 
can display microcausality in some suitable classical limit; that is, explaining how gauge invariant  
operators (Dirac observables) at spacelike distances can commute in generically curved spacetimes.
There it was demonstrated that a discretized relational framework approach to on shell background independent gauge theories of gravity
can possess an emergent light cone structure and microcausality provided there exist finite ranged nonlocal interactions,
which do not have to be of Planck scale. 
This allows classical spacetime to acquire microcausality naturally even if the underlying quantum geometry does not possess that property.
However, those results were based on Lieb-Robinson methods adopted from solid-state  physics \cite{Lieb}
and did not provide a microscopic origin for the classical spacetime nonlocalities. Here that gap is filled by examining \emph{commutatively} deformed
4-dimensional \emph{curved} Lorentzian  manifolds where the nonlocal action of the (3+1)-diffeomorphism symmetries is described by 
Hopf algebras possessing a suitable Drinfeld twist. 
The deformed diffeomorphisms' nonlocal action on the physical fields differs from the pointwise action in undeformed classical general relativity, 
and those deformed symmetries obey a distinct  Lie algebra, implying different physics.
Aside from the commutative $\star$-product, it is sufficient to consider Hopf algebras with twists
satisfying an Abelian constraint on their vector field generators \cite{Schenkel}\cite{AC_Abel}. 
Imposing background independence requires those generators to be self-consistently related to matter fields. As a result
it is found that the subtly deformed, but still classical, theory of spacetime naturally produces a nonlocality length $\xi_c$
which can be larger than the Planck length $L_P$,
and so spacetime acquires microcausality at longer lengths via the Lieb-Robinson route \cite{PdV}. Spacetime remains
classical in the sense that its gravitational degrees of freedom are not promoted to operators and their quantum fluctuations are ignored,
while the matter fields on spacetime are quantum. 
Unexpectedly, the requisite matter fields (particle zoo) may be
dark matter candidates. In some cases they have global continuous and discrete symmetries, and may even display symmetry breaking condensate ground states. 
In a sense using a Drinfeld twist to produce an on shell nonlocality length  $\xi_c$ is complementary 
to the asymptotic safety point of view \cite{Ambjorn}\cite{Reuter}, where $\xi_c$ emerges as a correlation length associated
with a renormalization group fixed point.  It is also noteworthy that theories of gravitation based on nonlocal vector bosons and second
rank tensors were explored starting in the early 1990's.\cite{Moffat} \\

The outline of the remainder of the article is as follows: Section 2 provides a hopefully self-contained introduction to the basics of Hopf
algebras and deformed differentiable manifolds. This overview relies heavily on \cite{Schenkel}, and
is not intended to be complete, but is rather  directed towards general readers in order to equip them with some intuition regarding Hopf algebras.
Section 3 discusses the physically relevant particular case of commutative 4-dimensional curved Lorentzian manifolds and Abelian twists,
and the technical problems it overcomes. 
Section 4 imposes the requirement of background independence on such twists by introducing a new sector of matter fields, and their associated 
continuous and discrete symmetries are then examined.  It also presents the 
idea of self-consistent twisting.
Section 5 provides estimates for the nonlocality length $\xi_c$ implied by this physical picture and leads the reader on a guided tour of the particle zoo. 
The theoretical and experimental relationship with the standard model is discussed.
Section 6 examines how the new particles could be dark matter candidates.
Section 7 turns to deformed classical electromagnetism:
The first order deformed Maxwell equations are derived, and the dispersion and attenuation of plane waves traversing cosmological distances through a dilute homogeneous gas of the new particles are discussed.
Section 8 concludes the article with a self-criticism of the approach, followed by possible directions for future work, and a brief summary.\\

\section{Introduction to Hopf algebra methods}

Why Hopf algebras? General relativity (GR) is a diffeomorphism (diff) invariant theory.   A diff may either be viewed as active, that is as a transformation of all the fields by dragging them to new coordinates keeping 
some arbitrary set of coordinate frames fixed (an alibi); or it may equivalently be viewed passively, as a field transformation under a coordinate transformation (an alias). 
Diff symmetry distinguishes GR from the other
symmetries in the standard model on flat (Minkowskian) spacetime. In this sense GR is a fully background independent theory:  the coordinates have no physical meaning, and
are merely calculational bookkeeping devices; there are no special coordinate points.  Instead one speaks of events, e.g. this is the event where particles A and B interact
and transform into C and D. It is the fields at events and their relationships that describe physical reality; distance and geometry themselves arise from these fields.\cite{Rovelli}
In GR, \emph{infinitesimal} diffs act in an event-wise fashion. The word ``point" should really be interpreted as ``event."
That is, the infintesimal diffs in GR depend only on what is going on at an event (and its infinitesimal neighborhood), so they are ultralocal 
and their action on objects occurs over a vanishing proper distance. We are interested in deformed GR, 
where the action of infinitesimal diffs becomes nonlocal over some proper distance $\xi_c\ne 0$. 
Hopf algebraic methods describe this utilizing a coordinate-free language, thereby maintaining the essential background independence of GR.  This requires a deformed concept
of tensors (which are defined by their transformation rule under diffs), as well as deformed Levi-Civita connections and covariant derivatives. Then a deformed Riemann curvature
tensor can be defined, and one can finally write down a deformed action for the geometric degrees of freedom (gravity). Moreover, it is found that a special class of Hopf algebras encapsulates the key notions of monotonicity and braiding, which are necessary to keep
gauge theories of canonical gravity from violating gauge invariance; i.e., they should be anomaly free.
If deformed classical manifolds were merely a matter of replacing pointwise products with a nonlocal ones, then there would be no need to use the abstract technology
of Hopf algebras. However, if one is concerned about the role of symmetries, their Lie algebras, and background independence, Hopf algebras are necessary because they describe how those
symmetries act on objects living in spacetime while keeping the theory mathematically consistent.  \\
 
The story starts out with an $D$-dimensional differentiable manifold $\mathcal{M}$, and the vector space of smooth real or complex 
vector fields $\Xi$ on $\mathcal{M}$.  This vector space is equipped 
with the $\mathbb{C}$-bilinear anti-symmetric Lie bracket $[u, v]$ for $u,v \in\Xi$, which obeys the Jacobi identity. The Lie algebra $L=(\Xi , [\cdot , \cdot])$ may be interpreted as infinitesimal
diffs of $\mathcal{M}$, which drag fields an infinitesimal distance along the local value of the vector field.  
The action of $L$ on objects such as functions, vectors, differential forms, or tensors residing on $\mathcal{M}$ 
is given via the Lie derivative $\mathscr{L}_{v}$
satisfying $\mathscr{L}_{v} \circ \mathscr{L}_{w} - \mathscr{L}_{w} \circ \mathscr{L}_{v} = \mathscr{L}_{[v, w]}$. Here $\circ$ denotes composition of operations.
Notice that a Lie derivative along a vector field is a background independent operation: acting on a tensor(density) it produces a tensor of the same type (and weight).
Not all derivations do this, for example the coordinate derivative acting on a tensor generally does \emph{not} yield a tensor.
Also associated with the infinitesimal diff $v$ is its inverse $-v$. 
One can also define a unit $1$ of $L$ by setting $\mathscr{L}_{1}\doteq \mathrm{id}$. Moreover $\mathscr{L}_{v}$ acts on tensor products of tensors or fields $\tau ,\tau '$ 
according to the Leibniz rule: $\mathscr{L}_{v} (\tau \otimes\tau ') = \mathscr{L}_{v}(\tau ) \otimes\tau ' + \tau\otimes\mathscr{L}_{v}(\tau ')$. One may also introduce a normalization 
map $\epsilon:\Xi\rightarrow\mathbb{R}$ called the co-unit such that $\epsilon(v) = 0, \forall v\in\Xi$ and $\epsilon (1) =1$.  From $L$ one constructs its universal enveloping algebra
$U\Xi$ as follows: First let $\mathcal{A}_{\mathrm{free}}$, the free associative and unital algebra, be the set of all finite sums of finite products of vector fields and $1$. 
The Lie bracket information is encapsulated in the ideal $\mathcal{I}$ freely generated by the elements $u v -v u -[u,v], \forall u,v\in\Xi$, where product is denoted by juxtaposition. $U\Xi$
is then defined as the factor algebra $\mathcal{A}_{\mathrm{free}}/\mathcal{I}$. The elements of $U\Xi$ act on objects inhabiting $\mathcal{M}$ from the left via the representation
$\mathscr{L}_{\xi\eta} =\mathscr{L}_{\xi}\circ\mathscr{L}_{\eta}$ and $\mathscr{L}_{1} = \mathrm{id}$. The product on $U\Xi$: $\mu : U\Xi\otimes U\Xi\to U\Xi$ is also sometimes denoted
by juxtaposition: $\mu (\xi\otimes\eta) = \xi\eta$. When dealing with these objects and operations it is vital to clearly distinguish the tensor product $\otimes$ from the 
juxtaposition products (composition of Lie derivatives).  Why introduce $U\Xi$ at all? The reason is that it describes the action of an arbitrarily large number of Lie derivatives, which is what 
distinguishes nonlocality from ultralocality. 
The Leibniz rule tells one how the algebra (of such arbitrarily large number of Lie derivatives) acts on tensor products of objects on $\mathcal{M}$. 
This is formally represented in the co-product map $\Delta : U\Xi\to U\Xi\otimes U\Xi$,
which acts on the generators $v\in\Xi$ of $U\Xi$ as $\Delta(v)=v\otimes 1 + 1\otimes v$ and $\Delta(1)=1\otimes 1$. This is then extended to all all $\xi\in U\Xi$ by $\mathbb{C}$-linearity 
and multiplicativity, i.e. $\Delta(a\xi + b\eta) = a\Delta(\xi) + b\Delta(\eta)$ for $a,b\in\mathbb{C}$, and $\Delta(\xi\eta)= \Delta(\xi)\Delta(\eta)$. It is common to introduce the sumless 
Sweedler notation: $\Delta(\xi) = \xi_{1}\otimes\xi_{2}$ with an implied summation. If $m\in M$ and $n\in N$ where $M,N$ are vector spaces acted on by $U\Xi$, 
one has $\xi(m\otimes n) \doteq \Delta(\xi) (m\otimes n) = (\xi_{1}\otimes\xi_{2})(m\otimes n) = (\xi_{1} m)\otimes(\xi_{2} n)$. 
These definitions preserve the Lie bracket structure: $\Delta(uv-vu) = \Delta([u,v])$. The inversion map $v\to -v$ is extended from $\Xi$ to $U\Xi$ by introducing the antipode $S$
(analogous to group element inversion) so that $S(v) =-v$ and $S(1)=1$. Then $S$ is defined on all of $U\Xi$ by $\mathbb{C}$-linearity and anti-multiplicativity: $S(\xi\eta) = S(\eta)
S(\xi)$, consistent with one's intuition from inverses of successive mappings. The co-unit  is likewise extended to all of $U\Xi$ by linearity and multiplicativity: $\epsilon(\xi\eta)=
\epsilon(\xi)\epsilon(\eta)$. \\

For the quintuple $H\doteq(U\Xi, \mu, \Delta,\epsilon, S)$ to be promoted to become a Hopf algebra, three further conditions must be imposed. Using sumless Sweedler notation to write
$\Delta(\xi)=\xi_{1}\otimes\xi_{2}$, one requires
\begin{align} 
(\xi_{1_1} \otimes\xi_{1_2})\otimes\xi_{2} & = 
\xi_{1}\otimes(\xi_{2_1}\otimes\xi_{2_2}) , \label{a} \\
\epsilon(\xi_1)\xi_{2}& = \xi = \xi_1\epsilon(\xi_{2})  \label{b} \\ 
S(\xi_1)\xi_{2} & = \epsilon(\xi) 1 = \xi_{1}S(\xi_2), \forall\xi\in U\Xi. \label{c}
\end{align}
(\ref{a}) and (\ref{b}) taken together mean $H$ is a co-algebra over $\mathbb{C}$. (\ref{a}) and (\ref{b}) make the quintuple $H$ into both an associative
unital algebra as well as a co-associative algebra with a co-unit that are compatible in the following sense: Co-multiplication $\Delta$ and co-unit $\epsilon$ are both algebra homomorphisms, 
or equivalently, the multiplication $\mu$ and unit $1$ are both co-algebra homomorphisms.  \\

Given a Hopf algebra $H$ one may ask if the co-product $\Delta(\xi) = \xi_{1}\otimes\xi_{2}$ is co-commutative. This idea is analogous to commutativity of an algebra, where
$\mu(\xi\otimes\eta) =\xi\eta = \eta\xi= \mu(\eta\otimes\xi)$. Co-commutativity means the co-opposite co-product $\Delta^{\mathrm{co}}(\xi) \doteq \xi_{2}\otimes\xi_{1}$, with tensor product 
factors in inverted order, obeys $\Delta(\xi) = \Delta^{\mathrm{co}}(\xi)$. If so, $H$ is called a co-commutative Hopf algebra. Generally Hopf algebras are neither 
commutative nor co-commutative.
Even if $H$ is non-co-commutative, $\Delta^{\mathrm{co}}(\xi)$ equals $\Delta(\xi)$ up to conjugation by an element $R\in H\otimes H$ independent of $\xi$, called the universal R-matrix. \\

We will be interested in deformations of Hopf algebras. Physically this means we wish the non-infinitesimal diffs $\xi\in U\Xi$ to act on objects 
inhabiting the manifold $\mathcal{M}$ in a way that is different, or deformed, from the standard (ultra)local way.  
This is accomplished by introducing the Drinfeld twist, which is an invertible element $\mathcal{F}\in H\otimes H$. $\mathcal{F}$ must satisfy two conditions:
\begin{align}
\mathcal{F}_{12}\circ(\Delta\otimes \mathrm{id})\circ\mathcal{F} & = \mathcal{F}_{23}\circ(\mathrm{id}\otimes\Delta)\circ\mathcal{F},  \mathrm{and} \label{twist_con_1} \\
(\epsilon\otimes\mathrm{id})\circ\mathcal{F} & = 1 = (\mathrm{id}\otimes\epsilon)\circ\mathcal{F}, \label{twist_con_2}
\end{align}
with $\mathcal{F}_{12} \doteq \mathcal{F}\otimes 1$, $\mathcal{F}_{23}\doteq 1\otimes\mathcal{F}$. One has $\mathcal{F}=1\otimes1 +\mathscr{O}(\lambda)$, 
where $\lambda$ is a complex variable parametrizing the deformation, and the 
nonvanishing $0$-th order part of $\mathcal{F}$ is necessary and sufficient to assure its invertibility.
(\ref{twist_con_1}) guarantees that deformed products ($\star$-products) of scalar valued functions is associative: $f\star(g\star h) = (f\star g)\star h$. The second condition ensures
that $f\star 1= f = 1\star f$. It is common to decompose $\mathcal{F}$ in  Sweedler notation as $\mathcal{F}= f^{\alpha}\otimes f_{\alpha}$, and its inverse as $\mathcal{F}^{-1} = 
\bar{f}^{\alpha}\otimes\bar{f}_{\alpha}$, where the $f$ and $\bar{f}$ are all elements of $H$, i.e. elements of $U\Xi$. The twist $\mathcal{F}$ transforms $H$ into a new (deformed)
Hopf algebra $H^{\mathcal{F}} = (U\Xi, \mu, \Delta^{\mathcal{F}}, \epsilon, S^{\mathcal{F}})$ with $\Delta^{\mathcal{F}} (\xi) = \mathcal{F} \Delta(\xi) \mathcal{F}^{-1}$, $S^{\mathcal{F}}
=\chi S(\xi) \chi^{-1}$, with $\chi \doteq f^{\alpha} S(f_{\alpha})$ and $\chi ^{-1} = S(\bar{f}^{\alpha}) \bar{f}_{\alpha}$. It is straightforward to show that 
$(\Delta^{\mathcal{F}})^{\mathrm{co}} (\xi) = R \Delta^{\mathcal{F}} (\xi) R^{-1}$ where $R=\mathcal{F}_{21} \mathcal{F} \in H\otimes H$ and 
$\mathcal{F}_{21}\doteq f_{\alpha} \otimes f^{\alpha}$
is $\mathcal{F}$ with interchanged tensor product ``blocks'' or ``legs,'' and $\mathcal{F}_{21}^{-1} = \bar{f}_{\alpha} \otimes \bar{f}^{\alpha}$ is the corresponding quantity for $\mathcal{F}^{-1}$.
$R=R^{\alpha}\otimes R_{\alpha}$ is the universal R-matrix, with inverse $R^{-1} = \bar{R}^{\alpha}\otimes\bar{R}_{\alpha}$.
 A nontrivial R-matrix is the signature of a non-co-commutative Hopf algebra.  \\

If the R-matrix further satisfies 
\begin{align}
(\Delta^{\mathcal{F}} \otimes\mathrm{id}) R & = R_{13} R_{23}\;\;\;\mathrm{and}\label{QT_con_1} \\
(\mathrm{id}\otimes\Delta^{\mathcal{F}}) R & = R_{13} R_{12}, \label{QT_con_2}
\end{align}
where $R_{12}= R\otimes 1$, $R_{23}=1\otimes R$, $R_{13}=R^{\alpha}\otimes1\otimes R_{\alpha}$ and $R_{21}=R_{\alpha}\otimes R^{\alpha}$, then it is referred to as 
a quasi-triangular Hopf algebra. If in addition to (\ref{QT_con_1}) and (\ref{QT_con_2}), $R_{21}=R^{-1}$ it is called triangular. These conditions have important physical meanings.
A quasi-triangular Hopf algebra satisfies the Yang-Baxter equation $R_{12} R_{13} R_{23} = R_{23} R_{13} R_{12}$, and quasi-triangularity is preserved by the deformation $H\to
H^{\mathcal{F}}$ when $\mathcal{F}$ satisfies the twist conditions  (\ref{twist_con_1}) and (\ref{twist_con_2}) above. It turns out that if the Yang-Baxter equation is fulfilled,
and $V$ is a module over $H$,
then $V$ can be used to construct a representation of the braid group $B_n$, with $V^{\otimes n}$ as the carrier space. This means that a quasi-triangular Hopf algebra encodes the physical 
braid-like qualities of monotonicity
and absence of self-intersections. It is noteworthy that unitary gauge flow in the relational formalism of gauge theories of gravity was previously found to possess precisely 
these same features.\cite{PdV}
In that work, the presence of a synchronizing external ``time'' together with monotonicity and ``non-collisional channels'' were necessary conditions for the formalism to remain anomaly free. 
We will not recapitulate those details here, but instead we can say the requirement that a Hopf algebra be quasi-triangular may be interpreted as a necessary condition to ensure that the 
deformed (3+1)-diffs will keep the 
gauge theory mathematically self-consistent.  All the twists used in this article will be quasi-triangular.\\

To make matters more concrete, suppose $\mathcal{M}$ is an $D$-dimensional manifold with local coordinates $x^{\mu}$. If the derivatives $\partial_{\mu}$ provide a global
basis for the tangent bundle of $\mathcal{M}$, any $v\in\Xi$ may be expressed as $v=v^{\mu}\partial_{\mu}$. The so-called Moyal-Weyl  (MW) twist on Minkowski (flat) spacetime is given 
for $\lambda\in\mathbb{C}$ by
\begin{equation}
\mathcal{F}_{MW} = \exp \left(-i\,\frac{\lambda}{2} \,\theta^{\mu\nu}\,\partial_{\mu}\otimes\partial_{\nu}\right). \label{MW_twist}
\end{equation}
For real valued $\theta^{\mu\nu}$ constant over $\mathcal{M}$ and anti-symmetric in its indices $\mathcal{F}_{MW}$ 
satisfies the twist conditions (\ref{twist_con_1}) and (\ref{twist_con_2}), and 
is also quasi-triangular. Its nontrivial R-matrix is $R=\exp[i\lambda\, \theta^{\mu\nu}\,\partial_{\mu}\otimes\partial_{\nu}]$.
Because of its relative simplicity $\mathcal{F}_{MW}$ is often used in 
quantum field theories on noncommutative \emph{flat} $\mathbb{R}^D$. It can be slightly generalized to curved spacetimes by the so-called Abelian twist
\begin{equation}
\mathcal{F}_{\mathrm{Abel}} = \exp\left(-i \,\frac{\lambda}{2} \,\theta^{ab} X_a\otimes X_b\right), \label{AS_Abel_twist}
\end{equation}
where $X_{a}\in\Xi$ for $a\in[1,N]$ are a set of $N\le D$ pointwise linearly independent, mutually Lie commuting vector fields: 
\begin{equation}
[X_{a}, X_{b}] = 0. \label{Abelian_X}
\end{equation}
$\theta^{ab}$ is an constant real (numerical) anti-symmetric $N\times N$ matrix. Such an Abelian twist 
generally also  possesses a nontrivial R-matrix, and so is 
noncommutative. The Abelian condition assures that the $\star$-product will be associative.
 It is essential to note that for this to be a background independent twist, the vector fields $X_a$ must be interpreted as \emph{Lie} derivatives, a
garden variety derivation such as $\partial_{\mu}$ is not guaranteed to accomplish this. In this way the Abelian condition (\ref{Abelian_X}) 
is also generally covariant, since the Lie bracket of two vectors is a vector,
and the vanishing of a vector (or any tensor) is a generally covariant statement. 
Hence if the undeformed theory was background independent, as GR is, so is the deformed one.
There is a useful result due to 
P. Aschieri and  L. Castellani \cite{AC_Abel} that states the following: 
Define the regular manifold $\mathcal{M}_{\mathrm{reg}}$ to be the set of all regular points of $\mathcal{M}$. A point $p\in\mathcal{M}$ is 
regular if there is an open neighborhood of $p$ over which $\mathrm{dim}(\mathrm{span} \{ X_a \})$ is constant. $\mathcal{M}_{\mathrm{reg}}$ is dense in $\mathcal{M}$. 
Aschieri and Castellani 
demonstrated that if for each $p\in\mathcal{M}_{\mathrm{reg}}$ there is open region $U\subseteq\mathcal{M}_{\mathrm{reg}}$ where the $X_a$ are pointwise linearly independent, then for 
each $p\in\mathcal{M}_{\mathrm{reg}}$ (i.e. almost everywhere in $\mathcal{M}$) there is an open region $U_p\owns p$ where there is a local basis called a ``nice basis'' $\{e_a \in\Xi : 
a=1,\dots, D \}$ such that the vector fields $X_a$ generating the twist commute with all the $e_a$, and the $e_a$ all mutually commute: $[e_{a}, e_{b}]=0$. If an Abelian twist is generated by 
pointwise linearly independent vector fields, then it reduces \emph{locally} to the MW twist. This is similar to non-twisted curved spacetimes, which are \emph{locally} 
diffeomorphic to flat spacetime.
 Not all Abelian twists reduce locally to an MW twist, and these are called exotic twists. In this 
article we assume the  twists are all non-exotic Abelian, i.e. the $X_a$ are eventwise linearly independent and mutually Lie commuting.  
Non-exotic Abelian twists are mathematically decorous:
one can define deformed tensor fields, deformed
Levi-Civita connections and covariant derivatives locally on $\mathcal{M}_{\mathrm{reg}}$, which are then extended to all of $\mathcal{M}$ by smoothness. \\

Given a twist $\mathcal{F}$, we wish to use it to construct deformed scalars (functions), vectors, differential forms, and tensors on $\mathcal{M}$. 
Suppose we are given $h,k\in C^{\infty}(\mathcal{M})$. Their algebra arises from pointwise multiplication $\cdot$, which is covariant under the (undeformed) Hopf algebra $H$, namely 
$\xi({h\cdot k})\doteq\Delta(\xi)(h\cdot k) = \xi_{1}(h)\cdot \xi_{2} (k)$, where we have abbreviated the Lie derivative action $\mathscr{L}_{\xi}$ as simply $\xi$. Once deformed by $\mathcal{F}$, 
however, the standard pointwise product $\cdot$ is no longer covariant under $H^{\mathcal{F}}$, but functions from $C^{\infty}(\mathcal{M})$ can be made into an algebra covariant
under $H^{\mathcal{F}}$ by deforming the pointwise $\cdot$ into a nonlocal product of functions as
\begin{equation}
h\star k  \doteq \bar{f}^{\alpha}(h)\cdot\bar{f}_{\alpha}(k), \label{Star_pdt_defn_A} 
\end{equation}
that is
\begin{equation}
(h\star k) (x)  = \mathcal{F}^{-1} (h(x), k(x) ), \label{Star_pdt_defn_B}
\end{equation}
which satisfies associativity $h\star(k\star l) = (h\star k)\star l$ and is unital $h\star 1 =h=1\star h$, but is generally noncommutative. In fact,
\begin{equation}
h\star k = \bar{R}^{\alpha}(k) \star \bar{R}_{\alpha}(h). \label{R_commtn}
\end{equation}
Hopf algebras with trivial R-matrices will then have commutative $\star $-products. Note carefully the distinction between commutative $\star$-products 
and Abelian twists (made of Lie commuting vector field generators $X_a$).
By using the MW (\ref{MW_twist}) or Abelian (\ref{AS_Abel_twist}) twists above, and expanding the exponential in a power series, its easy to see that $(h\star k)(x)$ contains Lie
derivatives of arbitrarily high orders, and so is non(ultra)local.
By covariance under $H^{\mathcal{F}}$ is meant 
  \begin{align}
\xi(h\star k) & = \Delta^{\mathcal{F}}(\xi) (\bar{f}^{\alpha}(h) )\cdot(\bar{f}_{\alpha}(k) ) \nonumber \\
 & = [\xi_{1}(\bar{f}^{\alpha}(h) ) ]\cdot [\xi_{2}(\bar{f}_{\alpha}(k) ) ] \nonumber \\
 & = (\bar{f}^{\beta} f^{\gamma}\xi_{1}\bar{f}^{\alpha} ) (h)\,\cdot\,(\bar{f}_{\beta}f_{\gamma}\xi_{2}\bar{f}_{\alpha} ) (k) \nonumber \\
 & = \xi_{1\mathcal{F}}(h)\star \xi_{2\mathcal{F}}(k), \label{Defd_Covarnc} 
\end{align}
where we have inserted a factor $1\otimes 1 = \mathcal{F}^{-1}\circ\mathcal{F}=$ 
$(\bar{f}^{\beta}\otimes\bar{f}_{\beta}) \circ (f^{\gamma}\otimes f_{\gamma})$ 
$=(\bar{f}^{\beta}f^{\gamma})
\otimes(\bar{f}_{\beta}f_{\gamma})$. Hence we have the following intuition: while $H$ describes the action of diffs of $\mathcal{M}$ on functions from the algebra 
$C^{\infty}(\mathcal{M},\, \cdot\,)$ that 
uses the pointwise product $\cdot$, $H^{\mathcal{F}}$ describes the actions of deformed diffs of $\mathcal{M}$ on functions from the algebra $C^{\infty}(\mathcal{M},\, \star\,)$ 
which utilizes the nonlocal product $\star$. \\

This Hopf covariant deformation procedure can be extended to the exterior (differential) calculus of $n$-forms $(\Omega\doteq\bigoplus_{n=0} \Omega^{(n)}, \wedge,\mathrm{d})$
 with undeformed wedge product $\wedge$ and 
exterior derivative $\mathrm{d}$.  A similar method can be used to produce deformed contractions of vectors and forms, and products of functions and vectors,
which all become nonlocal operations, unlike their undeformed counterparts. 
As we will not be using these, the details are omitted.  However, since tensors are important in deformed GR, we briefly describe how those operate.
Let $(\mathcal{T},\otimes_{A})$ be the tensor algebra generated by  $\Omega^1$ and $\Xi$
with $\otimes_{A}$ being the tensor product over the algebra $A=(C^{\infty}(\mathcal{M}),\,\cdot\,)$. The deformed tensor product reads as
\begin{equation}
\tau\otimes_{A\star} \tau ' \doteq\bar{f}^{\alpha}(\tau)\otimes_{A}\bar{f}_{\alpha}(\tau '),
\end{equation}
for all $\tau ,\tau ' \in\mathcal{T}$. \\

We will be interested in commutative deformations where $f\star g =g\star f$. In what way, if any, is the deformed theory different from the undeformed theory? 
This concern arises from the Gel'fand-Naimark theorem, which states there is a (Gel'fand) isomorphism between any commutative C$^*$-algebra 
$\mathscr{A}$ (fields) 
with unit which is an isometric
$*$-isomorphism from $\mathscr{A}$ to the commutative algebra of continuous $\mathbb{C}$-valued functions with pointwise multiplication. That is, might a commutative
deformation be simply un-done by such a field deformation (isomorphism)? Here it is shown that for the non-trivial  $\mathscr{F}\ne 1\otimes 1$ commutative non-exotic Abelian
twists of interest here, the answer is no.\\

To this end, let us construct a Hopf algebra $H_{\star}$ isomorphic (denoted $\simeq$) to $H^{\mathcal{F}}$.
First, the undeformed Hopf algebra $H$ acts on itself via the adjoint action $\mathrm{Ad}_{\xi}(\eta) \doteq \xi_{1} \eta S(\xi_{2})$. One deforms this by defining a $\star$-product on $H$ as
\begin{equation}
\xi\star\eta\doteq\mathrm{Ad}_{\bar{f}^{\alpha}} (\xi) \mathrm{Ad}_{\bar{f}_{\alpha}} (\eta), \; \xi,\eta\in U\Xi.
\end{equation}
$(U\Xi, \mu_{\star})$ and $(U\Xi,\mu)$ turn out to be isomorphic  \emph {as algebras}  under the mapping $D(\xi) = \mathrm{Ad}_{\bar{f}^{\alpha}}(\xi) \bar{f}_{\alpha}$. Notice $D(\xi)$ is 
linear in $\xi$. Using the algebra isomorphism $D$, one pulls the Hopf algebra structure on $H^{\mathcal{F}}$ back
to a Hopf algebra structure on $(U\Xi,\mu_{\star})$ called $H_{\star}=$ $(U\Xi, \mu_{\star},\Delta_{\star}, \epsilon_{\star}, S_{\star})$ having $D(1)=1$, with the Hopf structures related by:
\begin{align}
\Delta_{\star} & = (D^{-1}\otimes  D^{-1})\circ \Delta^{\mathcal{F}}\circ D, \nonumber \\
\epsilon_{\star} & = \epsilon\circ D, \nonumber \\
S_{\star} & = D^{-1} \circ S^{\mathcal{F}}\circ D, \nonumber 
\end{align}
and R-matrix $R_{\star} = (D^{-1}\otimes  D^{-1}) (R)$, which is also quasi-triangular.
$H_{\star}$ acts on objects living on $\mathcal{M}$ by the $\star$-Lie derivative
\begin{equation}
\mathscr{L}^{\star}_{\xi} \doteq \mathscr{L}_{D(\xi)}.
\end{equation}
One has for $ v\in\Xi$, $\Delta_{\star}(v) = v\otimes1 + D^{-1}(\bar{R}^{\alpha})\otimes \bar{R}_{\alpha}(v)$ and $\mathscr{L}^{\star}$ obeys the deformed Leibniz rule:
\begin{equation}
\mathscr{L}^{\star}_{v}(\tau \otimes_{A\star} \tau ' ) = \mathscr{L}^{\star}_{v}(\tau)\otimes_{A\star}\tau ' + \bar{R}^{\alpha}(\tau)\otimes_{A\star} \mathscr{L}_{\bar{R}_{\alpha}(v)} (\tau '). \label{Defd_Leibniz}
\end{equation}
The vector fields $v,w,z\in\Xi$ then may be assigned the deformed Lie bracket 
\begin{equation}
[v,w]_{\star} \doteq [\bar{f}^{\alpha}(v), \bar{f}_{\alpha} (w)]. \label{Defd_Lie_brack}
\end{equation}
This obeys the deformed anti-symmetry and Jacobi relations
\begin{align}
[v,w]_{\star} & = - [\bar{R}^{\alpha}(w), \bar{R}_{\alpha}(v)]_{\star},  \label{Woro_1} \\
[v, [w, z]_{\star}]_{\star} & = [[v,w]_{\star} , z]_{\star} + [\bar{R}^{\alpha}(w), [\bar{R}_{\alpha}(v), z]_{\star} ]_{\star}. \label{Woro_2}
\end{align}
This structure defines a so-called Woronowicz quantum Lie algebra and describes the infinitesimal deformed diffs, analogous to 
how the original Lie algebra $L=(\Xi, [\cdot,\cdot])$ described the undeformed diffs.
That is, this quantum Lie algebra of $H^{\mathcal{F}}$ as a vector space is $D(\Xi)\subset U\Xi$; it contains arbitrarily higher
order products of the vector fields and so encodes the notion of non(ultra)locality. \\

Now specialize to the case where there is a commutative twist. Then the Woronowicz quantum Lie algebra of infinitesimal deformed diffs
is a genuine Lie algebra, as the R-matrices are all trivial (unity). So we have the Lie algebra of deformed diffs $[v,w]_{\star} = [\bar {f}^{\alpha}v, \bar{f}_{\alpha}w]$.
Using the definition of the Abelian twist (\ref{AS_Abel_twist}), one finds
\begin{equation}
[v,w]_{\star}-[v,w] = \sum_{n=1}^{\infty} (-i\lambda/2)^{n}\frac {1}{n!} (\theta_{a_{1}b_{1}}\cdots \theta_{a_{n}b_{n}} ) [X_{a_1} \cdots X_{a_n} \cdot v,
X_{b_1} \cdots X_{b_n} \cdot w]. \label{DfftLie}
\end{equation}
In a ``nice" basis $\{ e_b\}$, $[X_a, e_b]=0$ for all $a,b$, and $X\cdot v = (X \cdot v^{\mu}) e_{\mu} = (\mathscr{L}_{X} \cdot v^{\mu}) e_{\mu} \ne 0$,
for non constant $v$. So the right hand side of (\ref{DfftLie}) is in general non-vanishing.
Moreover, since the expression (\ref{DfftLie}) involves arbitrary powers of Lie derivatives 
(nonlocal action) with respect to the varying nonzero vector fields $X$,
these deformations are not merely a globally constant multiplicative rescaling (isomorphism) of the undeformed diffs' Lie bracket. At the same time, (\ref{DfftLie}) also shows $[v, w]_{\star}$ cannot be expressed in the background independent, locally rescaled form 
$f(v,w) [v,w]$, where $f$ is some local scalar function depending only on $v,w$ and possibly their finite order Lie derivatives. Hence, in spite of the existence of a function algebra isomorphism,
 the deformed $\lambda\ne 0$ infinitesimal diffs (vector fields) possess a  Lie algebra non-isomorphic
to the undeformed $\lambda=0$ ones, provided the $X$ are non-vanishing over some set of non-zero measure. In Section 4 below the $X_a$ will be given  as constant multiples of currents carried by the fields. Those currents are gauge invariant, conserved, and/or are associated with stable particles, 
so it will not be physically possible to set them to zero almost everywhere merely 
by a field redefinition.  The symmetries  of the deformed and undeformed theories are then indeed 
distinct, and thus so is the physics. Hence such deformations are not merely field redefinitions of  the pointwise product limit, they transform the symmetries as well.
In fact, this is precisely a Wigner-In\"on\"u contraction\cite{Inonu_Wigner}\cite{Weinberg_IW}  in the $\lambda\to 0$ limit, the local diffs being the contraction of the nonlocal diffs.  
Similar contractions reduce the Poincar\'e Lie algebra to the Galilei one in the low velocity non relativistic limit
and the Moyal bracket Lie algebra (equivalent to quantum commutators) to the Poisson Lie algebra in the classical limit $\hbar\to 0$. \\

To make this point more rigorously, we now explicitly prove that even though there is an isomorphism $m$ of the algebra of \emph{functions} (scalar fields) 
between $H$ and $H'\doteq H_{\mathcal{F}}\simeq H_{\star}$ by the Gel'fand-Naimark theorem,
there is no mapping of \emph{vectors} $v\in \Xi \to m(v) \in \Xi'\doteq\Xi_{\mathcal{F}}$ such that 
\begin{equation}
[v,w]_{\star} = [m(v), m(w)], \forall v,w\in \Xi . \label{ProveMe}
\end{equation} 
First we note that according to \cite{Aschieri_2}, $\Xi = \Xi_{\mathcal{F}}$ as vector spaces. 
Suppose there were such a mapping $m$ of vectors satisfying (\ref{ProveMe}), then one must have
for $\alpha, \beta\in\mathbb{C}$ that 
\begin{equation}
[\alpha u+\beta v,w]_{\star} = [m(\alpha u+\beta v), m(w)]. \label{PM_2}
\end{equation}
But the LHS of this is $\alpha[u,w]_{\star} + \beta[v,w]_{\star} = \alpha [m(u), m(w)] +\beta [m(v), m(w)] $ $= [\alpha m(u)+ \beta m(v), m(w)]$. 
Therefore (\ref{PM_2}) implies
\begin{equation}
[m(\alpha u + \beta v) - \alpha m(u) -\beta m(v), m(w)]=0,
\end{equation}
for all $\alpha,\beta\in \mathbb{C}$ and $u,v,w\in\Xi$. Consequently
\begin{equation}
m(\alpha u + \beta v) = \alpha m(u) + \beta m(v), \label{PM_Linear}
\end{equation}
and thus $m$ is a linear mapping on vectors from  $\Xi$ to $\Xi'=\Xi$. From this it also follows that $m(\vec{0}) =\vec{0}$ and $m(-v)=-m(v)$.
It is easy to show that if two Hopf algebras are isomorphic 
under such a \emph{linear} mapping of vectors $m$ as Hopf algebras $m:H\to H'$ 
(meaning that $m$ also preserves all of the co-product $\Delta$, antipode $S$, and co-unit $\epsilon$ structures), then $m$  also preserves the Lie algebra structure: $m([v,w]) = [m(v), m(w)]$. 
To see this,
the Lie algebra of a Hopf algebra may be extracted by defining vectors $v$ as those elements $v$ of $H$ satisfying $\Delta(v) = v\otimes 1 + 1\otimes v$ and $ [v,w] \doteq v w - w v$ using the juxtaposition product. So if $\Delta (\xi)=
\xi_{1}\otimes\xi_{2}$, then the linear Hopf algebra isomorphism $m$ implies $m(\Delta(\xi))= m(\xi_1\otimes\xi_2) = \Delta'(m(\xi)) = (m(\xi))_{1'}\otimes (m(\xi))_{2'}$, and $m(\Delta(v)) = m(v\otimes 1
+1\otimes v) = m(v)\otimes 1' + 1' \otimes m(v)$, where the Hopf algebra units $1,1'$ obey $1'=m(1)$. 
By the Hopf algebra isomorphism $m(\Delta(v)) = \Delta' (m(v))$, so vectors of $H$ are mapped to vectors of $H'$, and vice-versa. 
Also by linearity of $m$, $m([v,w]) = m(vw-wv)= m(v)m(w)-m(w)m(v)= [m(v),m(w)]$. 
So if a linear vector mapping $m$ of Hopf algebras does not preserve the Lie algebra structure of the vector space(s) $\Xi,\Xi'$, then
 $m$ is not a Hopf isomorphism, and at least one of the co-product, antipode, and/or  co-unit is not preserved by $m$.  
Now notice the (bijective) mapping $D$ used to define the deformed $H_{\star}\simeq H_{\mathcal{F}}$ from the undeformed $H$ does not preserve the Lie algebra 
structure on $\Xi$. Specifically, the definitions show that $D([v,w])\ne [D(v), D(w)]$, where as noted previously $D(v)$ is linear in $v$.  Hence $H_{\mathcal{F}}$ and $H$ 
have non-isomorphic Lie algebras, and so if there is an $m$ satisfying (\ref{ProveMe}), they cannot be isomorphic as Hopf algebras. Therefore at least one of the following three cases
must hold: \\

(a) The co-product is not preserved by linear $m$: $(m\otimes m)\Delta(v) \ne \Delta' (m(v))$. The LHS equals $ (m\otimes m)(v\otimes 1 + 1 \otimes v) = m(v) \otimes 1'+ 1'\otimes m(v)$.
The RHS is $m(v)\otimes 1' + 1'\otimes m(v)$. So (a) is contradiction if $m(v)\in\Xi'$. \\

(b) The antipode is not preserved by linear $m$: $m(S(v))\ne S'(m(v))$. Note the definition (\ref{c}) applied to $H$ implies $S(v)+v=\vec{0}$. Applying the linear mapping $m$ to this
gives $m(S(v)) + m(v) = \vec{0}$ or $m(S(v)) = -m(v)$. Then the condition (b) yields $S'(m(v)) \ne m(S(v)) = -m(v)$, and $m(v)\in \Xi' =\Xi$ does not satisfy (\ref{c}) for $H'$, also
a contradiction for $m(v) \in \Xi'$. \\

(c) The co-unit is not preserved by linear $m$: $m( \epsilon(v)) \ne \epsilon ' (m(v))$. Applying $m$ to $\epsilon(v)=0$, one finds $m(\epsilon(v))=m(0) =0$. But (c) then implies 
$\epsilon'(m(v)) \ne m(\epsilon(v)) =0$, once again a contradiction for $m(v) \in \Xi'$. \\

Hence the original assumption that there is a \emph{vector} mapping $m$ satisfying (\ref{ProveMe}) must be incorrect. 
This further implies that the action of commutatively deformed infinitesimal diffs $v\in\Xi$ 
considered here is not merely the undeformed action of ``morphed'' diffs $v'(x^{\mu}) = v(x^{\mu} + \Delta^{\mu}(x) )\in\Xi$, for some smooth vector ``displacement" fields $\Delta^{\mu}$. This is 
a more intuitive way of perceiving that the deformed $H_{\mathcal{F}}$ and undeformed $H$ are non-isomorphic Hopf algebras; even though they possess the same
infinitesimal diffs $v\in\Xi$ and there is a Gel'fand-Naimark isomorphism of their \emph{function} algebras, 
the action of the deformed and undeformed diffs and their symmetries \emph{are} indeed distinct.\\

A similar result was obtained by \cite{Lizzi} for commutative deformations on flat (translationally invariant) spacetime, 
where it was demonstrated that, although there is a field redefinition (function algebra isomorphism) relating to the undeformed case, 
there is no corresponding \emph {Hopf algebra or co-algebra} homomorphism preserving the 
co-product and antipode structures.\\  

\section{Technical issues and commutative deformed products}

Attempts to construct noncommutative field theories and noncommutative classical gravitation have foundered on several obstacles.
We discuss those here and explain how commutative and co-commutative Hopf algebras circumvent those issues. We now work in $D=3+1$
dimensions. \\

One well-known difficulty with the MW twist $\mathcal{F}_{MW} =\exp[-i(\lambda/2) \theta^{\mu\nu} \mathscr{L}_{\mu}\otimes \mathscr{L}_{\nu}]$ for anti-symmetric
and constant matrix $\theta$ is the lack of causality \cite{Soloviev} \cite{Szabo}. This problem arises generically when $\theta$ is invertible. The invertibility
permits some Gaussian integrations. This allows two scalar fields with overlapping supports of size $\delta\ll || \theta ||^{1/2}\doteq \xi$ to have a $\star$-product
with support having size $||\theta||/\delta \gg ||\theta||^{1/2} = \xi$. This may be circumvented by requiring $\theta^{\mu\nu}$ to have at least one vanishing
eigenvalue. This will obstruct the requisite Gaussian integrations. \\

Another technical issue in using the MW twist or its generalized Abelian cousin (\ref{AS_Abel_twist}) is that the important gravitational deformed Einstein and Riemann
tensors are no longer real valued in a generic spacetime \cite{Schenkel}. Replacing the imaginary anti-symmetric $i\theta^{\mu\nu}$ by a general Hermitian matrix suffers from the same problem. 
However, utilizing a real MW twist of the form
\begin{equation}
\mathcal{F}_{RMW} \doteq \exp \left( -\frac{\lambda}{2}\, \theta_{S}^{\mu\nu} \mathscr{L}_{\mu}\otimes\mathscr{L}_{\nu}\right) \label{F_RMW}
\end{equation}
where $\lambda\in\mathbb{R}$, $\theta_{S}$ is a real (numerical) symmetric matrix, and $\mathscr{L}_{\mu}$ denotes the Lie derivative with respect to the $\mu$-th basis vector $\hat{e}_{\mu}$, has no such reality difficulties. Specifically, twists of the MW or RMW type obey the complex conjugation relations
\begin{align}
\mathcal{F}_{MW}^{*\otimes *} & =(S\otimes S) \mathcal{F}_{MW\,21}, \;\;\;\mathrm{whereas} \\
\mathcal{F}_{RMW}^{*\otimes *} & =(S\otimes S) \mathcal{F}_{RMW\,21}=\mathcal{F}_{RMW}.  \label{Reality}
\end{align}
If one generalizes the real MW twist (\ref{F_RMW}) to an real Abelian twist of the form 
\begin{equation}
\mathcal{F}_{C} \doteq \exp \left( -\frac{\lambda}{2} \,\theta_{S}^{ab} X_{a}\otimes X_{b}\right), \label{F_C}
\end{equation}
with $[X_{a}, X_{b}]=0$ and pointwise linearly independent $X_{a}$, then it also obeys (\ref{Reality}), and then one can construct generally real valued deformed Riemann and Einstein
tensors necessary for a classical theory of gravity with deformed diffs.
The proof that $\mathcal{F}_{C}$ generates an associative $\star$-product is the same as  
original MW case.  Henceforth we will omit the subscript $S$ on $\theta^{ab}$ when we use $\mathcal{F}_{C}$. \\

The reader may be wondering about the reason for the subscript $C$ on $\mathcal{F}_{C}$. This is because for bosonic vector fields $X_a$ and bosonic functions $f,g$,
 the twist $\mathcal{F}_{C}$ generates a commutative $\star$-product,
$f\star g = g\star f$. For $f$ and $g$ both fermionic $f\star g = - g\star f$; and if one of $f, g$ is bosonic and the other fermonic, $f\star g = g\star f$. 
It might be thought possible to take each $X_a$ as 
fermionic (real Grassmann 4-vector fields), that would require real anti-symmetric $\theta^{ab}$ to make $f\star g = g\star f$ for bosonic $f,g$, but the Abelian constraint on $X_a$ would lead to
$X_{a} X_{b} = - X_{a} X_{b}=0$. It also would lead to a violation of the requirement that $f\star g = g\star f$ if one of $f,g$ is bosonic and the other fermionic. 
Hence Grassmann vector fields cannot
enter the commutative twist. To distinguish the $\star$-product generated by $\mathcal{F}_{C}$ from its generally 
noncommutative cousin, the $\star$-product, we designate the former commutative product
by the symbol $\odot$ instead of $\star$. \\

Another issue plaguing noncommutative classical general relativity is that while one can construct a $\star$-covariant derivative compatible with the $\star$-metric tensor, 
that derivative is not generally compatible with the $\star$-inverse metric \cite{Schenkel}. Here inverse is defined using $\star$-matrix multiplication. However for the commutative $\odot$-product,
this obstacle vanishes, and all the standard derivations of textbook tensor analysis then carry through. \\

There are also some technical issues with gauge symmetries in noncommutative field theory. Namely, they are based on a fiber bundle construction that places conditions on the transition 
functions (trivializations) on overlapping patches that involve $\star$-products of group-valued functions.\cite{Nakahara} For matrix gauge (structure) groups this means matrix multiplication, inversion, and 
determinants are  likewise computed by the $\star$-product. So while one can define closed $\star U(N)$ gauge groups, $\star SU(N)$ are no longer generally closed \cite{Szabo}. 
This conundrum also
disappears for the $\odot$-product. At the same time, the Seiberg-Witten (field redefinition) map \cite{Seiberg}\cite{Ulker}, utilized extensively in noncommutative field theory,
has a straightforward generalization to the $\odot$-product, 
which will be put to good use in a later section. \\

A vital symmetry for classical gravitation is local Lorentz invariance. A general Abelian twist will violate global Lorentz and rotational invariance under ``particle"  transforms
since there the ingredients of the twist ($\theta^{\mu\nu}$ or the $X_{a}$) are not transformed while the matter fields are \cite{Carroll}. 
This leads to predicted anisotropy effects in the cosmic microwave background 
for noncommutative field theories \cite{Akofor}. 
To restore Lorentz (and rotational) invariance one must relate the twist to the matter contents of the theory, so that the theory becomes background independent. That is,
if $\theta^{\mu\nu}$ or the $X_a$ are arbitrarily given or chosen properties of the manifold, then one is violating background independence. To regain it, $\theta^{\mu\nu}$ and the $X_a$ must 
come from somewhere physically. This is analogous to general relativity where spacetime curvature is related to the matter determined stress-energy tensor; at a spacetime event the 
curvature (a manifold property) must physically arise from some source. This is carried through for the commutative twist in a subsequent section. \\

The commutative twist $\mathcal{F}_{C}=f^{\alpha}\otimes f_{\alpha}$ possesses several further delightful simplifications that we now mention. By expanding the exponential in (\ref{F_C}) it is 
straightforward to demonstrate that $f^{\alpha} = f_{\alpha}$ and $\bar{f}^{\alpha} = \bar {f}_{\alpha}$, that is the ``blocks" of the twist's tensor product are balanced and real valued to all orders.
From $R=R^{\alpha}\otimes R_{\alpha}=\mathcal{F}_{21} \mathcal{F}^{-1}$ for the R-matrix, one readily derives $R^{\alpha} = R_{\alpha}=\bar{R}^{\alpha}=\bar{R}_{\alpha}=1$, 
as expected for a commutative (and co-commutative) twist, and was also used at the end of the previous section to show that a commutative twist cannot be undone by a field definition.
Another complication for noncommutative field theory, fortunately absent for $\mathcal{F}_{C}$, is that spinor and $\gamma$-matrix calculus becomes more elaborate \cite{AC_Spinor}.
This is transparent from examining the identity
\begin{equation}
[A(x)\gamma_{A}, B(x)\gamma_{B}]_{\star} = \left(1/2 \right) [A(x),B(x)]_{\star} \{\gamma_{A},\gamma_{B} \} + \left(1/2 \right) \{A(x),B(x)\}_{\star} [\gamma_{A},\gamma_{B} ].
\end{equation}

The $\odot$-product of objects on $\mathcal{M}$ induces a corresponding deformation (nonlocality) in the phase space underlying the relational framework.\cite{PdV}\cite{Dittrich}
In that approach, the canonical variables
are functions on a 3-dimensional slice $\Sigma\subset\mathcal{M}$ (a gauge fixing), namely $q^{i}(y)$ and $p_{i}(y)$, where $y$ coordinatizes $\Sigma$ 
and $i$ is a discrete index labeling the different gauge constraints. The scalar functions $f,g$ above are replaced by $q^{i}(y)$ and $p_{i}(y)$ on $\Sigma$ so only the parts 
of the twist generating vectors $X_a$ tangential to $\Sigma$ act on the canonical variables or other objects made from them. Specifically, the Poisson bracket of phase space functions $f,g$
is deformed to 
\begin{equation}
\{f, g\}^{PB}_{\odot} = \sum_{i} \int_{\Sigma} \mathrm{d}^{3}y \left( \frac{\delta f}{\delta q^{i}(y)} \odot \frac{\delta g}{\delta p_{i}(y)} -
                           \frac{\delta f}{\delta p_{i}(y)} \odot \frac{\delta g}{\delta q^{i}(y)} \right).
\end{equation}
This expression satisfies the requirements that a Poisson bracket must anti-commute, distribute over addition, follow the Leibniz product rule, and obey the Jacobi identity.
These in turn follow from using a trivial R-matrix in relations (\ref{Defd_Leibniz}) through (\ref{Woro_2}) describing the deformed (Woronowicz) Lie algebra for the $\odot$ product.
This way gauge flow in phase-space  will be deformed, which can equivalently be viewed as a deformation of the associated symplectic 2-form.
There are also deformations of the symplectic vector space used to describe the quantum scalar fields among other objects living on $\mathcal{M}$.\cite{Schenkel}
We will examine the length scales of the deformations below.\\
                            
To summarize: The commutative and cocommutative twists considered here may be perceived as less mathematically interesting than their 
more sophisticated noncommutative relatives from the Big City, however by circumventing several obstacles
they do allow one to extend general relativity and
field theory to deformed manifolds where nonlocality can play a role. And that's precisely what is needed physically.   \\

\section{Actions, twist-matter coupling, symmetries, and deformation self-consistency}

We are now ready to write down an action for deformed gravitation. We seek to describe classical spacetime as 
invariant under deformed diffs, in distinction to GR which is invariant with respect to 
undeformed diffs.  
Since the $\odot$-product is covariant
under deformed diffs,
we may replace the pointwise product in the undeformed Einstein-Hilbert action throughout with the $\odot$ product.  The deformed Riemann tensor 
and its various contractions are then similarly computed
from deformed Levi-Civita connections and the $\odot$-product. There are actually two forms of the Einstein-Hilbert action, one
(Palatini or first order) taking the the tetrad $e$ and spin connection $\omega$ to be independent variables, and the second order one where $\omega=\omega[e]$.\cite{Rovelli}
Since it has not been experimentally resolved which best corresponds to reality, we will use the simpler second order form. Similar procedures apply to the first order formulation
of GR. 
For instance the deformed (torsion-free) Einstein-Hilbert action with cosmological constant $\Lambda$ reads
\begin{align}
S^{\odot}_{EH} & = \frac{1}{2\kappa} \int_{\mathcal{M}} \mathrm{d}^{4}x \; | g^{\odot} |^{1/2} \odot (R^{\odot}-2\Lambda), \;\;\; \mathrm{with} \label{EH_Action} \\
\kappa & = 8\pi G/{c^4}.
\end{align}
There are no Gibbons-Hawking-York terms since we are taking $\partial \mathcal{M} =\emptyset$ for simplicity.
The classical deformed gravitational field equations then become
\begin{align}
& R^{\odot}_{\mu\nu} - \frac{1}{2}  R^{\odot} \odot g^{\odot}_{\mu\nu}  +\Lambda g^{\odot}_{\mu\nu} + \Delta_{\mu\nu}(\lambda)  
= \kappa \, T^{\odot}_{\mu\nu} \;\;\; \mathrm{with} \label{Einstein_eqns} \\
& T^{\odot}_{\mu\nu}  = -2 \frac{\delta \mathcal{L}^{\odot}_{M} }{\delta g_{\odot}^{\mu\nu} } + g^{\odot}_{\mu\nu} \odot\mathcal{L}_M^{\odot}. 
\label{Stress_energy} 
\end{align}
The last term on the left hand side of equation (\ref{Einstein_eqns}) arises from the dependence of  $\odot$ on the inverse metric tensor, and is
has leading order $\mathscr{O}(\lambda^1)$.
In equation  (\ref{Stress_energy}) $\mathcal{L}^{\odot}_{M}$ is the deformed Lagrangian density for \emph{all} matter (non-metric tensor) fields, 
including the  twist producing matter fields.
$T_{\mu\nu}$ still obeys a deformed version of energy momentum conservation. Regarding the matter and Yang-Mills actions, those may also be rendered invariant under deformed diffs by replacing the pointwise multiplication by $\odot$ throughout.
Hence the classical electromagnetic (Maxwell) action is 
\begin{align}
S^{\odot}_{EM} & = -(1/4) \int_{\mathcal{M} }\mathrm{d}^{4}x \; | g^{\odot} |^{1/2} \odot F^{\odot}_{\mu\nu} \odot F_{\odot}^{\mu\nu}, \label{EM_Action} \\
F^{\odot}_{\mu\nu} & = \nabla ^{\odot} _{\nu} A^{\odot}_{\mu} -  \nabla ^{\odot} _{\mu} A^{\odot}_{\nu}= A^{\odot}_{\mu,\nu} - A^{\odot}_{\nu,\mu},
\end{align}
where $\nabla^{\odot}_{\mu}$ denotes the deformed spacetime covariant derivative.  The general Yang-Mills action is also easy to write down, but since we will not be requiring it,
nor the deformed QCD or electroweak actions, and will not display them here. Also we will henceforth drop the $\odot$ 
designation for deformed quantities when that meaning is clear from the context, and similarly for the $C$ on $\mathcal{F}_{C}$. \\

We now turn to $\mathcal{F}$ and discuss how it is constructed from matter fields, and investigate the properties of the pointwise linearly independent and 
Abelian generating vectors $X_{a}, a=1,\dots,N$. 
We may think of the $X_a$ as spanning an $N$-dimensional, $a,b$-indexed internal space, which we will refer 
to as GM space, after Groenewold and Moyal who pioneered the $\star$-product.
As discussed in the previous section, to ensure causality on a $D=3+1$ Lorentzian manifold, 
the eigenvectors of $\theta^{\mu\nu} = \theta^{ab} X_{a}^{\mu}
X_{b}^{\nu}$ with nonzero eigenvalues must span a space of dimension no greater than 3. Hence $1\le N\le 3$. Similarly there are $DN(N-1)/2$ independent real differential
constraints of the form $[X_{a},X_{b}]=0$ with $DN$ degrees of freedom in the set of $X_a$. This implies $(N-1)/2\le 1$ or $1\le N\le 3$, 
the same as the causality constraint. Since the Lie bracket
of two vector fields is a vector field, the Abelian constraint is automatically a covariant statement. One might wonder
 whether the vanishing of the deformed Lie bracket might be more appropriate.
However, from the definition (\ref{Defd_Lie_brack}) of the deformed Lie bracket and equation (\ref{DfftLie}), it is easy to work out that the two conditions are equivalent. To enforce the Abelian constraint one could
include a suitable Lagrange multiplier term in the total action, however it turns out to be simpler to construct Abelian $X_a$ directly,
as will be shown shortly. Also, the Abelian constraint should not be viewed as a gauge constraint, but rather as a technical condition necessary for the theory to maintain strict mathematical
propriety, like possessing an associative product. Finally, we note the Latin indices $a,b$ on $\theta^{ab}$ (labeling which vector) may be raised or lowered with impunity, since those indices live in a space whose metric is the Kronecker-delta. This is distinct from the Greek coordinate indices $\mu , \nu$ labeling the coordinate components of a given vector, 
which are lowered (raised)
by the (inverse of) the deformed spacetime metric tensor $g_{\mu\nu}$. 
For example, Lorentz transforms mix the Greek indices but not the Latin ones.\\

The $X_{a}$ may be composed from either familiar Standard Model (SM) fields
or non-SM matter fields. We will refer to the latter as GM fields or GM matter.
The procedure is: First construct \emph{one} $X$ vector field from matter. Next, classicize it. Finally,
construct any necessary additional  classical $X_a$ from the first one. We now perform this step by step.\\

The twist will insert arbitrarily many factors of the form
\begin{equation}
\Delta\mathcal{L}=-(\lambda/2) \theta^{ab} X^{\sigma}_{a} X^{\rho}_{b} (\mathscr{L}_{\sigma}\otimes\mathscr{L}_{\rho}). \label{F_Insertion}
\end{equation}
into the undeformed action. 
By well known arguments \cite{Weinberg2GInvar},   
maintaining the the overall gauge invariance of the deformed Lagrangian density under any gauge symmetries of the undeformed theory 
(such as SM gauges) requires
 that the standard Lie derivative $\mathscr{L}_{\mu}$ must be then be replaced by the gauge covariant Lie derivative
\begin{equation}
\hat{\mathscr{L}}_{\mu} \Psi_{l} = \mathscr{L}_{\mu} \Psi_{l} - i A^{\alpha}_{\mu} (t_{\alpha})^{\,m}_{l} \odot \Psi_{m} \label{GCovLie}
\end{equation} 
when acting from (\ref{F_Insertion}) on \emph{any} set of fields $\Psi_{l}$ appearing in the undeformed Lagrangian density 
that transform under some non-singlet 
representation of a gauge group with Lie algebra generators
$t_{\alpha}$ in the presence of the gauge potential $A^{\alpha}_{\mu}$. Gauge coupling constants (charges) are absorbed into the $t_{\alpha}$.
This occurs separately from constructing the $X_a$ from matter fields. 
When acting on gauge invariant objects like $U(1)$ field strengths $F_{\mu\nu}$,
the second term on the RHS is absent from that $\hat{\mathscr{L}}_{\mu}$ within $\odot$, but other cases generally require it. \\

Since we wish the overall deformed action
to have the same symmetries under charge conjugation (now including any possible GM charges) as well as spatial parity, that means the factor
(\ref{F_Insertion}) must be even under both those symmetries separately. In particular, it implies that $X$ must be either even or odd under charge conjugation, and similarly for parity.
This places strong restrictions on the admissible forms for $X$. For example, let us na\"ively try to construct one of the $X$ from some matter Dirac spinor field $\psi$ as:
\begin{equation}
X_{\mu} \stackrel {?}{=} \bar{\psi}\odot \mathscr{D}_{\mu} \psi \doteq \bar{\psi}\odot \partial_{\mu} \psi + \bar{\psi} \odot (-iqA_{\mu})\odot \psi + 
\bar{\psi}\odot\omega ^{IJ}_{\mu} \odot \Sigma_{IJ}\,\psi,
\end{equation}
where $\omega$ is the spin connection field, $I,J$ are (tangent space) Lorentz indices, and $\Sigma^{IJ} = (-i/4) [\gamma^{I}, \gamma^{J}]$. In the second order formalism
$\omega_{\mu}^{IJ} = e^{I}_{\nu}\odot \nabla_{\mu} \, e^{J\nu}$, where $(e_{\mu}^{I})$ $ e^{\mu}_{I}$ is the (co-) tetrad, and $\nabla_{\mu}$ is the (deformed) spacetime covariant derivative.
The first two terms come from the gauge covariant derivative, which for simplicity has been chosen to be a $U(1)$ gauge field 
with coupling constant $q$. The last term is the gravitational coupling, as the tetrad field reduces to a Kronecker delta in flat spacetime.
 While all three terms are of even spatial parity,
the first two are C-conjugation even, but the gravitational term is C-conjugation odd (because of $\Sigma$). Consequently, this choice of $X$
has an ill-defined overall sign under C-conjugation, and therefore is inadmissible.  The easiest way to avoid such problems is to start from the matter Lagrangian
 density $\mathscr{L}_{\mathrm{M}} (\psi)$. 
Then there are four general ways to construct the $X_a$ from matter, up to an overall constant of proportionality that may be absorbed into the scale of the twist $\lambda$, see equation
(\ref{F_C}). It turns out to be easiest to understand them by first assuming the matter that
produces the $X_a$ interacts by some set of gauge potentials $A^{\alpha}_{\mu}$.  We assume the matter gauge Lie groups are all compact
and have finite dimensional representations, and consequently the Lie generators may be taken to be Hermitian matrices. We  keep our mind  open to the possibility that the matter fields comprising the $X_a$ 
could couple to either some SM gauge fields $A_{\mu}$
or to gauge fields inhabiting only the GM sector.  We will examine alternative ways to construct the $X_a$ later on.
Then one may easily calculate the gauge derived current from those gauges by
\begin{align}
J^{\mu}_{\alpha} & \doteq \frac {\delta S[\psi]}{\delta A_{\mu}^{\alpha} } \\
& = (-i) \frac{\partial\mathscr{L}_{\mathrm{M}} } {\partial(D_{\mu} \psi)_l} \odot (t_{\alpha})^{\, m}_{l} \psi_{m} \\
& = i \sum_{l,m} \bar{\psi}_{m} \odot\gamma^{I} e^{\mu}_{I} (t_{\alpha})^{\,l}_{m} \odot\psi_{l},
\end{align} 
where $l, m$ are particle species indices, and  the matter Lagrangian density for fermions is
\begin{align}
\mathscr{L}_{\mathrm{M} }(\psi) & = -\sum_{l} \bar{\psi}_{l}\odot \left(\gamma^{I}e^{\mu}_{I}\odot (D_{\mu} \psi)_{l} + m_{l}\,\psi_{l} \right), \;\;\mathrm{with} 
\label{Fermion_Lagrangian} \\
(D_{\mu} \psi)_{l} & \doteq \nabla_{\mu} \psi_{l} -iA^{\alpha}_{\mu}(t_\alpha)_{l}^{\, m} \odot \psi_{m}, \;\;\mathrm{and}\\
\nabla_{\mu} \psi & \doteq \partial_{\mu} \psi +  \omega_{\mu}^{IJ}\odot \Sigma_{IJ}\,\psi,
\end{align}
where $\nabla_{\mu} \psi$ is the spacetime covariant derivative of the Dirac spinor $\psi$.
One way to compute \emph{one} of the real valued $X_a$ is to equate it to some gauge current (now dropping the GM subscript $a$ for clarity)
\begin{align}
X_{\alpha}^{\mu} & \doteq \binom{\mathrm{Re}} {\mathrm{Im}} J^{\mu}_{\alpha},\;\;\mathrm{with} \label{Gauge_X} \\
J^{\mu}_{\alpha} & = i \sum_{l,m} \bar{\psi}_{m}\odot \gamma^{I}e^{\mu}_{I}\odot(t_{\alpha})^{\,l}_{m} \psi^{l}.
\end{align}
This current is C-conjugation odd. 
In the case of nonrelativistic fermions, the imaginary part is preferred as it gives a vanishing time-component to the current 
when using the Dirac basis for $\gamma^{I}$. 
To ensure the insertions of the form (\ref{F_Insertion}) in the deformed action effected by the twist produce an overall gauge invariant deformed 
action starting from a gauge invariant undeformed action means
those insertions  must be gauge invariant, hence the $X_a$ must also be gauge invariant.
Under an infinitesimal gauge transformation parametrized by the real valued $\epsilon^{\alpha} (x)$
the matter fields transform as
\begin{align} 
\delta\psi_{l} & = i \epsilon^{\alpha} (x) (t_{\alpha})^{\,m}_{l} \odot \psi_{m}(x), \\
\delta (D_{\mu} \psi)_{l} & = i\epsilon^{\alpha} (x) (t_{\alpha})^{\,m}_{l} \odot (D_{\mu} \psi)_{m} (x),
\end{align}
and one finds
\begin{equation}
\delta X_{\beta}^{\mu} = \binom{\mathrm{Re}} {\mathrm{Im}} \sum_{l,m, \alpha} \epsilon^{\alpha}\odot \left( \bar{\psi}_{l}\odot \gamma^{I} e^{\mu}_{I}\odot\psi_{m}\right)
[t_{\beta}, t_{\alpha}]^{m}_{l}, \label{dXGauge}
\end{equation}
where the Hermiticity of the $t_{\alpha}$ has been utilized. 
We take the overall gauge Lie algebra to be a direct sum of commuting compact simple and $U(1)$ subalgebras.  
This is equivalent to requiring positive definiteness for the quantum mechanical inner product; 
that is, it ensures the absence of negatively normed states.\cite{Weinberg_SimpleLie}
Since direct sums of simple Lie algebras are semi-simple, they have no invariant subalgebras whose generators all commute with each other. 
So for some fixed non-$U(1)$ generator $t_{\beta}$, there will be at least one other generator $t_{\alpha}$ from $\beta$'s algebra with which 
$t_{\beta}$ 
does not commute. Consequently the RHS of equation (\ref{dXGauge}) vanishes only if  $t_{\beta}$ comes from a $U(1)$ subalgebra  (so proportional to
the identity matrix). Hence $X$ computed from equation (\ref{Gauge_X}) will be gauge invariant, $\delta X_{\beta} = 0$, 
only if $\beta$ is a $U(1)$ gauge, and then 
$(t_{\beta})^{\, l}_{m} =q_{l}\, \delta ^{l}_{m}$, $q_l$ being the dimensionless $U(1)$ charge. Hence the $U(1)$ current
\begin{equation}
X^{\mu}=  \binom{\mathrm{Re}} {\mathrm{Im}} \left(\sum_{l} iq_{l}\, \bar{\psi}^{l}\odot \gamma^{I}e^{\mu}_{I}\odot\psi^{l}\right) \label{XFC}
\end{equation}
is the \emph{only} gauge invariant 4-vector one can construct from gauges. By construction it is a conserved current: its deformed covariant divergence vanishes.
This way of constructing the first $X$ will be referred to as the $X\gamma\psi$ model.
The presence of $\odot$ in $X$ maintains the deformed diff invariance of the deformed action, and the implied self-consistency will be examined below.
Even though there may be non-Abelian $A^{\alpha}$ with corresponding non-vanishing 
gauge coupling constants, they cannot comprise
a gauge invariant \emph{vector} twist generator. If one turns off the $U(1)$ gauge by taking $q_{l} \to 0$ for all $l$, there is no more gauge based twist. But there could be non-gauge based 
 $X$ constructed in a similar way: Suppose $q_l$ were replaced by a (possibly species $l$ dependent) quantity $\mathscr{Q}_l$, 
which is also a $\psi_l$-field independent and gauge 
 invariant coordinate scalar, with well defined signs under each of $C,P, T$, then the
 corresponding $X$ comprised as in equation (\ref{XFC})  also inherits those desiderata. 
 One very simple possibility would be to set $\mathscr{Q}_l=1$, democratically independent of (fermion) species $l$. This will be referred to as the gaugeless or numerical model. 
 A species dependent alternative would be to set $\mathscr{Q}_l=m_l$, called a mass model. Finally, $\mathscr{Q}_l$ could be set to some species dependent quantum number,
 such as baryon number $B$ or electronic lepton number $L_e$, a quantum number model. 
  \\

Turning now to building one $X$ from scalar field(s),
 we take the multiplet of (complex) scalars $\phi_{l}$ to be described by the $\mathcal{F}_{C}$ deformation of the well known $\phi^4$ action in the presence of gauge fields:
\begin{equation}
S[\phi]= \sum_{l} \int_{\mathcal{M}} \mathrm{d}^{4}x \;| g |^{1/2} \odot \left[- (D^{\mu} \phi)_{l} \odot (D_{\mu} \phi)_{l}^{\dagger} -  m_{l}^2 \, \phi_{l}\odot \phi_{l}^{\dagger} - (g/2)
 (\phi_{l}\odot\phi^{\dagger}_{l})^{\odot 2} \right].  \label{Phi_4}
\end{equation}
Here $(D_{\mu}\phi)_{l}= \partial_{\mu}\phi_{l}-iA_{\mu}^{\alpha} (t_{\alpha})_{l}^{\, m}\odot\phi_{m}$ is the gauge (and spacetime) covariant derivative.
Then by construction, $S[\phi]$ is  gauge invariant, and so will be the entire action, provided any SM-GM interactions are made of
 suitable factors of $\phi_l$ together 
with the gauge covariant derivatives of the $\phi_l$, and all factors get spot welded together by $\odot$.
Computing the gauge currents and then using the same arguments as in the fermionic case one finds the conserved current,
\begin{equation}
J^{\mu}_{\beta} (\phi) = -i (D_{\mu} \,\phi)_{l}^{\dagger} \odot (t_{\beta})^{\, m}_{l} \phi_{m},
\end{equation}
implying
\begin{equation}
X^{\mu} =\binom{\mathrm{Re}} {\mathrm{Im}} \sum_{l} i q_{l}\,\phi_{l}\odot ( D_{\mu} \phi)_{l}^{\dagger},
 \label{XDPhi}
\end{equation}
for $U(1)$ charges $q_l$, which may then be likewise extended to the $\mathscr{Q}_{l}$. 
This $X$ is also C-odd, as well as conserved and gauge invariant for Hermitian Lie algebra generators.
This will be referred to as the $XD\phi$ model.
Like the fermionic $X$ in equation (\ref{XFC}), this is proportional to the $q_l$, but this scalar expression for $X$
depends on all the gauge fields with which the $\phi_l$ interact through $D_{\mu}$. 
The corresponding fermionic twist generator in equation (\ref{XFC}) has no such gauge field dependence. 
We will study the implications of these models as well as  alternatives to scalars and Dirac fermions in 
the next section, where numerical estimates together with symmetries will 
be used to further constrain the possibilities. \\

This is good place to mention that the $X_a$ cannot be any gauge boson field(s). That would make $\mathcal{F}$ gauge non-invariant,
regardless of whether or not some Englert-Brout-Higgs-Guralnik-Hagen-Kibble mechanism (hereafter, Higgs mechanism) has endowed the gauge field with mass.  
One might then be tempted to try $\theta^{\mu\nu} = F^{\mu\nu}$, for some gauge field strength tensor. However, that would be $\mu ,\nu$ antisymmetric;  commutative twists, however,
require a symmetric $\theta^{\mu\nu}$.
Hence photons, gluons, and the massive weak gauge bosons do not produce a twist.\\

Next one classicizes the $X_a$. Since the vectors $X_a$ generating the twist $\mathcal{F}$ are constructed from matter fields, it is important to understand
 when these fields are to be considered as quantum fields
and when they become classical, either as expectation values with respect to some state or as stationary points of an action.  Inside the non-$\mathcal{F}$ 
terms in the Lagrangian density (outside $\odot$), the fields like $\phi$ and the GM gauge potentials $A$ have their
 standard meanings, i.e. quantum or classical as one may freely choose, it is only inside $\mathcal{F}$ that more care is necessary. 
 Since we are building a model of classical spacetime, which
 $\mathcal{F}$ partially describes, the $X_a$ fields there must be given a classical interpretation so that there are no quantum fluctuations in the twist. One way to do this is by taking the 
 $X_a$ inside $\mathcal{F}$ to be comprised from fields such as $\phi$ or $\psi$ at the stationary points of their action. 
 The factor containing the tensor product of (gauge covariant) Lie derivatives acts to insert arbitrary powers of that differential operator 
 into the Lagrangian wherever $\odot$ occurs, prefixed by the same power of the classical factor $(\lambda/2)\theta^{ab} X_{a}\otimes X_{b}$. 
In this case the mass dimension of the $X_a$
 inside the twist will be its classical (canonical) dimension, as there is no quantum field renormalization of a classical field. 
As an alternative, one could construct the classical $X_a$ from
expectation values of the currents introduced earlier. Aside from questions regarding which matter state to use to evaluate the expectation value, one would then also have to be cautious 
about anomalous dimensions  potentially entering any dimensional analysis at the energy scales of the fields and coupling constants entering the twist.\cite{Weinberg_AD} 
In flat spacetime there is strong lattice gauge evidence that the $\phi^4$ theory is asymptotically free. That is, its matrix 
elements receive multiplicative corrections proportional to $1+\mathscr{O}(1/\ln E)$ at an energy scale $E$; and then the anomalous dimensions are zero since the corrections are powers of 
$\ln E$. Similarly for a $SU(N)$ gauge theory with $n_f$ fermionic (quark-like) flavors, the 1-loop beta function is $\beta_{1} = (g^{4}/16\pi^3) [-(11/6)N +(1/3)n_{f}]$,  
so a $SO(3)\simeq SU(2)$ gauged scalar theory
 is asymptotically free $\beta_{1} < 0$ for $n_{f}< 11$. Moreover, within the asymptotic safety scenario the running coupling constants 
 $G_E$ and $\Lambda_E$  are responsible for the non-Gaussian (non-trivial) fixed points, thereby at least allowing the scalars to be asymptotically free in curved spacetime.  \\
 
 Finally, now that we have just \emph{one} classical, matter dependent twist vector generator $X_1$ in hand, 
 how do we generate the $N-1$ others so that they are Abelianized: $[X_{a}, X_{b}]=0$ for all
 distinct pairs of $a,b$?  We first specifically address the $N=2$ case. 
Notice if one has obtained an Abelianized pair in one coordinate frame, the 
 transformed pair of fields will also be Abelian in any other frame because the vanishing of a vector field is a covariant statement.  Choosing some frame, the Abelian condition is equivalent to 
 the following system 
 of 4 first order linear partial differential equations for the components  of $X_{2}$:
 \begin{equation}
 (X_{1}^{\mu}\partial_{\mu}) X_{2}^{\nu} = X_{2}^{\mu} (\partial_{\mu} X_{1}^{\nu}).   \label{Abel_12}
 \end{equation}
 Provided $X_1 \ne0$,  this can be transformed into 
 \begin{equation}
 \partial_{\rho} X_{2} ^{\nu} = \sum_{\sigma \ne \rho} A^{\sigma} \partial_{\sigma} X_{2}^{\nu} + B^{\nu}_{\rho\mu} X_{2}^{\mu}, \label{CK_eqn}
 \end{equation}
 to which the Cauchy-Kovaleskaya theorem may be applied in a neighborhood when the (non-tensors) $A^{\sigma}$ and $B^{\nu}_{\rho\mu}$ are analytic functions of the coordinates through their dependence on $X_1$. Alternatively, one may use the fact that the Lie derivative can be expressed in identical form either with coordinate derivatives throughout or covariant derivatives in the absence of torsion, so in a ``nice" basis
 one obtains (\ref{Abel_12}) with $\nabla_{\mu}$ replacing $\partial_{\mu}$ everywhere. The theorem above states the system (\ref{CK_eqn}) then has a unique analytic solution in that neighborhood, given boundary values for $X_{2}^{\nu}$. 
 So then if there is some 3-slice of spacetime on which
 one has $[X_1, X_2]=0$, then one may integrate the solution off the slice.
 In particular since $X_1$ is constructed from the currents carried by a massive particle, it will be a timelike vector. Here timelike and spacelike are defined with respect to the deformed metric tensor (see self-consistent twisting below). Suppose one specifies some analytic data for $X_1$ on a spacelike hypersurface $\Sigma$. Choosing coordinates so that $\Sigma$ is a constant $x^0$, the PDE (\ref{Abel_12}) on $\Sigma$ where $X_{1}^{k}=0$ reads
 \begin{align}
(\partial_{0} X_{1}^{0}) X_{2}^{0} + (\partial_{j} X_{1}^{0}) X_{2}^{j} -X_{1}^{0} (\partial_{0} X_{2}^{0}) & =0, \;\;\mathrm{and} \nonumber \\
 (\partial_0 X_{1}^{k}) X_{2}^{0} -   X_{1}^{0} (\partial_{0} X_{2}^{k}) & = 0. \label{Sigma_PDE}
 \end{align}
From these one sees that the $X_1$ data together with the PDE (\ref{Abel_12}) and initial values for $X_2$ on $\Sigma$ determine the normal derivatives
of $X_2$ there. That is, the spacelike surface $\Sigma$ is a non-characteristic surface for the PDE. 
Alternatively, the PDE (\ref{Abel_12}) has the standard first order form 
 \begin{equation}
 \left(A^{\nu}(X_1)\right)^{\rho}_{\sigma} \frac {\partial u^{\sigma}}{\partial x^{\nu}} - B^{\rho} (u, X_1) =0, \label{MyPr}
 \end{equation}
with unknown $u^{\sigma} \doteq X_{2}^{\sigma}$ and $(A^{\nu}(X_1))^{\rho}_{\sigma} = X_{1}^{\nu}\, \delta^{\rho}_{\sigma}$.
Suppose the implicit form for $\Sigma$ is $\varphi=\varphi(x^{0},\ldots ,x^{3})=0$. The characteristic form of the PDE (\ref{MyPr})
 is defined as
 \begin{equation}
 Q(\partial\varphi/\partial x^{0},\ldots , \partial\varphi/\partial x^{0})  \doteq \det _{\sigma,\rho} \left[ \sum _{\nu=0}^{3} (A_{\nu})^{\rho}_{\sigma} \frac{\partial\varphi}{\partial x^{\nu}}\right],  \\
 \end{equation}
 which reduces here to $ Q= [X_1^{\nu}(\partial\varphi /\partial x^{\nu})]^{4}$. A characteristic surface is one for which $Q$ vanishes. 
To test if the PDE is hyperbolic,
consider the solutions of $Q(\lambda\xi+\eta) =0$ for the unknown scalar $\lambda$ with $\eta$ a spacelike
  vector tangent to $\Sigma$ at some point, and $\xi$ a timelike vector normal to $\Sigma$ at the same point. The solution is $\lambda = - \eta\cdot X_{1}/\xi\cdot X_{1}$, where the denominator
  is strictly negative for metric signature $(-+++)$. Hence $\lambda$ is a 4-fold real root, the Abelian PDE system 
  has any spacelike hypersurface $\Sigma$ as a non-characteristic surface, and the PDE system (\ref{Abel_12}) is (non-strictly) hyperbolic.
 Consequently, given any data on a spacelike $\Sigma$ for $X_{2}$, that data and the PDE system determines its normal derivatives. For example taking data to satisfy (\ref{Sigma_PDE})
  on $\Sigma$, one has a well-posed unique solution for the ``twin" $X_{2}$, provided the analyticity assumptions hold. 
 Hence to integrate off the spacelike slice requires those analyticity conditions to continue to hold, and once one has
 chosen the values of $X_2$ on $\Sigma$, the twin ``evolves"  according to (\ref{CK_eqn}), but is dependent on the matter field inside $X_1$. Places where $X_1$ vanishes
 are defects of some kind.  They would be a deformed version of Cauchy horizons, but are not as severe as the curvature singularities arising in classical GR: they signal a partial
 local breakdown of the
 predictability of the theory, but not a full blown divergence. 
 Since $X_1$ is a classicized current derived from the matter field $\phi$ or $\psi$, it may be possible to avoid its vanishing at isolated points where the quantum matter field
 is zero by smearing or taking the currents' expectation values. 
 But defects  where $X_{1}=0$ (which may not necessarily be isolated) are still physically possible and are intriguing twist-gravitational 
 analogs of vortices in superconductors or textural defects in the superfluid phases of $^3$He. Also note that because $X_1$ is generated
in a gauge invariant way from (the generalizations of) equations (\ref{XFC}) or (\ref{XDPhi}) above, 
so its twin $X_2$ is also gauge invariant by the preceding Cauchy-Kovaleskaya analysis,
and therefore the Abelian constraint is trivially preserved by matter gauge transformations.  This analysis of
 $[X_a, X_b]=0$ is also unaffected by the choice of the standard Lie derivative or its gauge covariant version, 
see equation (\ref{GCovLie}), also because the $X_a$ are gauge invariant.\\

Turning now to $N=3$, this case is readily demonstrated to be over-constrained. 
Suppose one already has one pair of twins $X_1$ and $X_2$ with $[X_{1}, X_{2}]=0$, and wishes to construct a 
third real valued $X_3$ that Lie commutes with the first two. That would generically impose a system of 8 independent 
PDEs on $X_3$, which only has 4 degrees of freedom at each spacetime event. 
That ends the story for $N=3$, unless one restricts the matter and gauge degrees of freedom.\\

The commutative twist $\mathcal{F} = \exp[-(\lambda/2)\theta^{ab} X_{a}\otimes X_{b} ]$ displays interesting symmetries.
 If there are $N$ linearly independent $X_{a}$, then $\mathcal{F}$ 
is invariant under global $O(N)$ symmetry; i.e. $N\times N$ real orthogonal linear transformations. It is easily verified that for the transformation 
\begin{align}
X_{a} \to \bar{X}_{a} & \doteq O_{ab} X_{b}, \\
\theta^{ab}\to \bar{\theta}^{ab} & \doteq (O\theta O)^{ab},
\end{align}
with $O\in O(N)$, that $\bar{\theta}^{ab}$ is also a real symmetric matrix. 
We may picture this global transform as a rigid rotation of the $X_a$ in GM space. However, this transform is not yet a symmetry of the action 
because the coupling constants $\theta^{ab}$ are also transformed together with the $X_a$. Since $\theta_{ab}$ is real and $a,b$ symmetric,
it always possible to find some basis by a real orthogonal transform so that the transformed $\bar{\theta}_{ab}$ is real diagonal in $a,b$.
There are only two cases of physical interest: $N=1$ and $N=2$. The former has a twist in the form $\mathcal{F} =\exp\left[-\,(\lambda/2) \,
\theta\, X_M\otimes X_M \right]$, $O(1)$ is trivial, and there is no global continuous symmetry of $\mathcal{F}$. For $N=2$ after diagonalization
the twist takes the form 
\begin{equation}
\mathcal{F}=\exp\left( -\frac{\lambda}{2} \theta\, \left[X'_{M}\otimes X'_{M} \pm k^{2} X'_{T}\otimes X'_{T} \right] \right),
\end{equation}
for $k, \theta$ nonzero real constants. 
Since the Abelian constraint $[X'_M, X'_T]=0$ is preserved under a global rescaling  $X'_M\mapsto\tilde{X}_T = k X'_T$ and 
$X'_M\mapsto\tilde{X}_M = X'_M$,
the $N=2$ twist can always be cast into the form
\begin{equation}
\mathcal{F}=\exp\left( -\frac{\lambda}{2} \theta [\tilde{X}_{M}\otimes \tilde{X}_{M} \pm \tilde{X}_{T}\otimes \tilde{X}_{T} ] \right). 
\label{NiceN_2}
\end{equation}
The sign is determined by the relative signs of the two real eigenvalues of the $2\times 2$ numerical matrix of coupling constants $\theta^{ab}$.
For the upper choice of sign, this form of the twist is invariant under the global transform of only the vector generators given by
\begin{equation*}
\begin{pmatrix}
\tilde{X}_M (x) \label {UpperGlbl}\\
\tilde{X}_T  (x)
\end{pmatrix}
\mapsto
\begin{pmatrix}
\cos\varphi & \sin\varphi \\
-\sin\varphi & \cos\varphi
\end{pmatrix}
\begin{pmatrix}
\tilde{X}_M (x)  \\
\tilde{X}_T (x)
\end{pmatrix}.
\end{equation*}
Obviously this $ 2\times 2$ matrix is an element of the familiar connected compact Lie group $\mathrm{SO}(2)$.
Similarly for the lower choice of sign one has the global symmetry
\begin{equation*}
\begin{pmatrix}
\tilde{X}_M (x) \label{LowerGlbl}\\
\tilde{X}_T  (x)
\end{pmatrix}
\mapsto
\begin{pmatrix}
\cosh\varphi & \sinh\varphi \\
\sinh\varphi & \cosh\varphi
\end{pmatrix}
\begin{pmatrix}
\tilde{X}_M (x)  \\
\tilde{X}_T (x)
\end{pmatrix}.
\end{equation*}
This matrix is an element of the real symplectic group $\mathrm{Sp}(2, \mathbb{R})$, a connected noncompact Lie group of dimension one.
 \\

Can this $N=2$ global symmetry be promoted to a gauge symmetry? 
The role of $\mathcal{F}$ in the action lies inside $\odot$, 
creating insertions into the Lagrangian density of arbitrary powers of (\ref{F_Insertion}).
For these insertions to be gauge invariant one requires \cite{Weinberg2GInvar}
\begin{equation}
\frac{\partial(\Delta\mathcal{L})} {\partial (X^{\mu}_{a}) } (t_{\alpha})_{a}^{\; b} \odot X^{\mu}_{b} =0, \label{GInvar_A}
\end{equation}
where the $t_{\alpha}$ are the $N\times N$ real anti-symmetric generators of the (undeformed) Lie algebra $\mathfrak{o}(N)$. It is straightforward to
verify the corresponding condition 
\begin{equation}
(t_{\alpha})^{a}_{c} [\theta ^{ab} X_{c}\otimes X_{b} + \theta^{ba} X_{b}\otimes X_{c} ] =0.
\end{equation}
Hence the global $O(N)$ can be gauged inside the insertions
introduced into the action by $\mathcal{F}$. \\

But this is not the end of the $N=2$ gauge story! What about the Abelian constraint $[X_M, X_T]=0$, is that $O(2)$ gauge invariant? 
Specifically, the spacetime position dependent  transformation in GM space modifies $X_M$
to some $X'_M$. 
If $N=2$, one then constructs the twin $X_T$ of $X_M$ as discussed earlier from $[X_M , X_T]=0$, 
and $X_T$ is also altered by the same  spacetime position dependent transformation in GM space to $X'_T$. Is $[X'_M, X'_T]=0$
for a general smoothly position dependent $SO(2)$ transformation of $X_M$ and $X_T$?  This issue does not arise for $N=1$ where there really is no 
meaningful transformation. So consider the following position dependent $SO(2)$ transform of the two twist generators
 for the upper choice of sign in equation (\ref{NiceN_2}):
\begin{equation*}
\begin{pmatrix}
X'_M (x) \\
X'_T  (x)
\end{pmatrix}
=
\begin{pmatrix}
\cos\varphi(x) & \sin\varphi(x) \\
-\sin\varphi(x) & \cos\varphi(x)
\end{pmatrix} \odot
\begin{pmatrix}
X_M (x)  \\
X_T (x)
\end{pmatrix},
\end{equation*}
parametrized by the now position dependent $\varphi(x)$. Using $[X_M , X_T]=0$, it is easy to show that
\begin{equation}
\left[X'_M, X'_T\right]^{\mu} = -\left(X_{M}^{\mu}\odot X_{M}^{\sigma} + X_{T}^{\mu}\odot X_{T}^{\sigma} \right) \odot \partial_{\sigma} \,\varphi(x),
\end{equation}
and similarly for the lower sign case in equation (\ref{NiceN_2}):
\begin{equation}
\left[X'_M, X'_T\right]^{\mu}= \left(X_{M}^{\mu}\odot X_{M}^{\sigma} - X_{T}^{\mu}\odot X_{T}^{\sigma} \right) \odot \partial_{\sigma} \,\varphi(x).
\end{equation}
Both of these are nonvanishing for generic $X_M$ unless $\varphi$ is position independent; that is, 
only \emph{global} $O(2)$ symmetries generally preserve the $N=2$ Abelian constraint, so the symmetry cannot be gauged.\\

For $N=2$ if we assume that the undeformed non-GM sector is invariant under global GM $O(2)$ of $X_M$ and $X_T$,
then the entire action is globally GM $O(2)$ invariant. This will be the case if the $X_a$ only enter the action
through the twist and the $\odot$-product.
Associated with this global symmetry there are Noether charges and their conserved currents. It is interesting to 
inquire a bit further into these Noether charges and currents.
Therefore consider the infinitesimal $O(2)$ transformation of the $X_a$ given by
\begin{align}
X_M & \mapsto X_M + \epsilon X_T , \\
X_T & \mapsto X_T - \epsilon X_M ,
\end{align}
where $\epsilon$ is an infinitesimal real constant.
The corresponding Noether current is then
\begin{equation}
J^{\mu}_{N} = \frac{\partial \mathscr{L} } {\partial (\mathscr{L}_{\mu} X_{M}^{\nu})} X_{T}^{\nu} -
\frac{\partial \mathscr{L} } {\partial (\mathscr{L}_{\mu} X_{T}^{\nu})} X_{M}^{\nu},
\end{equation}
where $\mathscr{L}$ is the deformed Lagrangian density and $\mathscr{L}_{\mu}$ is the standard Lie derivative.
This requires the evaluation of the quantity
\begin{equation}
\frac{\partial}{\partial (\mathscr{L}_{\mu} X_{a}^{\nu})} \left( X_{1} \cdots X_{n} \cdot F\right),
\end{equation}
with $F$ some factor appearing before or after an $\odot$-product inside $\mathscr{L}$. $F$ is not $X_a$ dependent by assumption. However,
because $[X_a ,X_b ]$ vanishes, one can commute the $X_{1} \cdots X_{n}$ so that the $X$ being acted
upon by ${\partial}/{\partial (\mathscr{L}_{\mu} X_{a}^{\nu}) }$ lies to the extreme left, and then the 
partial derivative operation gives zero. Hence the global Noether currents $J^{\mu}$ all vanish, as do the associated conserved charges. 
Alternatively put, for either value of $N$, both fermionic $X_a$ from (\ref{XFC}) and scalar based twist vector generators $X_a$ from (\ref{XDPhi}) carry no charges
 themselves, while the matter that constitutes them can.
So the Abelian constraint that maintains associativity of the $\odot$-product removes physical predictive power from Noether's theorem in this case,
as well as obstructing the promotion of global $O(2)$ symmetry to a $N=2$ gauge symmetry. 
In particular, global $O(2)$ GM symmetry does not place any constraints on GM matter's gauge groups or
 gauge coupling constants (GM charges).\\
 
The actions (\ref{Fermion_Lagrangian}) and (\ref{Phi_4}) together with $\mathcal{F}_{C}$ are additionally  invariant under the discrete $Z_2$ symmetry $\hat{\sigma}:\phi_a\to - \phi_a$
that acts as the identity on SM fields.
If this also applies to the SM-GM interactions, such as 
might be described by a $\phi$-Higgs interaction like $\mathcal{L} \propto (H^{\dagger}\odot H) \odot \phi_{a}\odot\phi_{a}$, 
and also if the ground state $|0\rangle$ is $Z_2$ symmetric $\sigma|0\rangle = 
|0\rangle$, so there is no GM ground state condensate, then this symmetry can protect against energetically allowed processes
 with an odd number of $\phi$ scalars decaying
 into purely SM products.
This may be easily seen by examining the quantity 
\begin{equation}
Q= \langle 0|(\prod_{n\in SM} \hat{a}_{n}) \,\hat{H}_{\mathrm{int}} \,\hat{a}^{\dagger}_{\phi} |0\rangle ,
\end{equation}
where $\hat{H}_{\mathrm{int}}$ is some SM-GM interaction, or the corresponding part of the (delta function free piece of) the S-matrix.
 All decay rates are proportional to $|Q|^2$. By inserting a factor $1=\hat{\sigma}^{-1}\hat{\sigma}$ after $\hat{H}_{\mathrm{int}}$ and using $[\hat{H}_{\mathrm{int}}, \hat{\sigma}] =0$, one finds
\begin{align}
Q & = \langle 0|(\prod_{n\in SM} \hat{a}_{n})\,\hat{\sigma}^{-1} \,\hat{H}_{\mathrm{int}} \,\hat{\sigma} \,\hat{a}^{\dagger}_{\phi} |0\rangle \nonumber \\
 & = -\langle 0|(\prod_{n\in SM} \hat{a}_{n}) \,\hat{H}_{\mathrm{int}} \,\hat{a}^{\dagger}_{\phi} |0\rangle=-Q,
 \end{align}
since $\hat{\sigma}$ and its inverse have no action on SM fields, and $\hat{\sigma}\hat{a}^{\dagger}_{\phi} |0\rangle = - \hat{a}^{\dagger}_{\phi} |0\rangle$. Hence $Q=0$, and the $Z_2$
 symmetry protects a single isolated $\phi$ particle from decaying into only SM particles under the above assumptions. This is stability is desirable for $\phi_a$, whose job requirements include
managing the ubiquitous $\odot$-product. \\
 
Aside from how GM matter might interact with SM fields, how are the theoretical foundations
of the SM affected by the GM matter generated commutative twist?
The SM gauge symmetries would all remain intact, since as discussed
in Section $3$, their fiber bundle structure still functions with the deformed commutative product and the same gauge groups. Internal
symmetries, like weak isospin, are also maintained since they are not acted on by diffs.
Similarly,  discrete symmetries are either unaffected by the twist's Lie derivatives or, 
for C, P, T and various combinations thereof, are preserved since the twist is C, P, T invariant. Lorentz invariance is also untouched since the twist has been explicitly constructed to maintain it, and its $SO(3,1)$ gauge structure remains in place. The SM quantum numbers are unchanged from the undeformed case. The effect of the twist on renormalizability and experimental measurements will be discussed in the next section, after estimates have been made for its size. \\

What happens to the cornerstones of classical gravitation \cite{Will}, namely the Equivalence Principles?
Commutatively deformed classical spacetime is not purely described by the metric tensor but also by the twist, which varies from event to event with its
 generators, the $X_a$.  The twist is then a geometric property of spacetime, describing how the (3+1)-diffs act nonlocally on all objects living in $\mathcal{M}$,
 including on the $\phi_a$ themselves, as may be seen from the presence of $\odot$ in (\ref{Phi_4}).
 However, a ``point'' mass (specifically meaning having a size sufficiently small that gravitational tidal forces are negligible) interacting purely gravitationally will still have a classical trajectory given by a 
 geodesic of the deformed metric tensor, so the weak Equivalence Principle
 holds. But the strong Equivalence Principle would be violated since spacetime is no longer solely described by the metric tensor, or by an equivalence class of metric tensors.
 Roughly speaking, the $\phi_a$ resemble the 
 Brans-Dicke theory's scalar dilation field $\Phi$ that describes variations of the Newton-Cavendish ``constant" $G$. Discussion of the Einstein Equivalence Principle, intermediate between
 the strong and weak, is hampered by semantics over whether the $X_a$ should be deemed ``matter" or ``geometry."  In a sense they are both. The fields like  the $\phi_{a}$ 
 and $\psi$ introduced above
 will act as a typical matter source of curvature via stress-energy is the usual way: Their Lagrangians (\ref{Fermion_Lagrangian}) and (\ref{Phi_4})generate a contribution
  to the stress-energy tensor (\ref{Stress_energy}) and then act as sources of the gravitational field by the gravitational field equations (\ref{Einstein_eqns}), and at the same time
   the twist produced by the $X_{a}(\phi)$ 
encodes how (3+1)-diffs operate.  So commutatively deformed general relativity, like Brans-Dicke theory, blurs the sharp distinction between matter and geometry that is a familiar feature of  Einsteinian gravity.  From this we also see the GM fields interact with each other and the SM fields gravitationally. \\
 
  The twist $\mathcal{F}$ and its matter field generators $X_{a}$ are computed self-consistently 
together with the all the matter and gauge fields.  This is necessary because both the twist generators $X_a$ as well as the
action of the gauge covariant Lie derivative $\hat{\mathscr{L}}_{\mu}$ on matter fields depend on the twist itself, as seen simply 
from the presence of the $\odot$-product in expressions (\ref{GCovLie}),(\ref{XFC}), and (\ref{XDPhi}). That is, the twist partially depends on itself.
A similar back-action or feedback situation also occurs in classical (and semi-classical) undeformed general relativity,
 since the gravitational fields there
are determined from the matter energy-momentum tensor, which in turn depends on the (quantum) matter dynamics,
 that is partially dependent on the gravitational fields, also requiring self-consistency for a solution.
 Now the twist gets caught up in this loop as well.
We sketch this procedure here, but emphasize from the start that it is schematic and formal at this stage, without
proof of its convergence.  For purposes of simplicity it is illustrated below using scalar GM matter.  Back-action effects 
are ignored in subsequent sections.\\
 
 We will apply the Seiberg-Witten map \cite{Seiberg} from the undeformed $(\phi,A)$ to the deformed $(\hat{\phi}, \hat{A})$ gauge theories.
Given some undeformed scalars $\phi_{a}^{[0]}$ and gauge potentials $A^{[0]}$ as ``seed fields," we will apply a commutative version of the Seiberg-Witten map
 and its corresponding differential equation to calculate the deformed fields for a fixed twist, 
and then iterate this process by computing a new twist from those deformed fields.\\
  
 In slightly more detail, the stages of the calculation are: \\

(0) Initialization: Start with the undeformed  (``seed") stationary points of $(\phi_{a}^{[0]}, A^{[0]})$, and compute $X_{a}$ from equation (\ref{XDPhi})
using the undeformed $A^{[0]}$ while setting the $\odot$ product to the ordinary product for this initialization.
Then $\mathcal{F}^{[0]} = \mathcal{F}(X^{[0]})$ is the seed twist. Set $n=0$.
 
(1) Keeping the twist fixed as $\mathcal{F}^{[n]}$ by freezing only the $X^{[n]}_a$ inside it,  deform the undeformed fields to arbitrary order in $\lambda$ with that fixed twist by 
acting on \emph{the seed} $( \phi^{[0]}, A^{[0]})$ using the Seiberg-Witten map 
recursion relations described below. This gives the deformed fields $(\phi^{[n+1]}, A^{[n+1]}) \doteq (\hat{\phi}, \hat{A})$ to all orders in $\lambda$ for  $\mathcal{F}^{[n]}$.

(2) Use the relations (\ref{XFC}) or (\ref{XDPhi}) to compute one deformed vector generator $X^{[n+1]}$ from $(\phi^{[n+1]},$ $ A^{[n+1]})$ 
using $\mathcal{F}^{[n]}$ to evaluate $\odot$ on the RHS (followed by classicizing it and constructing its twin, if necessary).
With those new generators,  calculate a new twist $\mathcal{F}^{[n+1]}$, and then return to (1) after incrementing $n$.

(3) Hope that Nature smiles gracefully, and that for ``sufficiently small deformations,''  She allows this ``deformation flow" to converge to a fixed point 
by some quantitative measure as $n$ increases or gets large. This deformation flow is reminiscent of renormalization group flow, except here the $\phi$ and $A$ fields
are flowing (together with all the other fields as well), but not the twist coupling constant $\lambda$.
Some order of magnitude estimates for ``sufficiently small deformations" will be presented later to justify this, and the relative size of the first order
deformation flow in the generators $X_a$ is found below in (\ref{FracX}) to be
 of order $10^{-28}$. \\

Employing a fixed twist $\mathcal{F}^{[n]}$
 at each depth of iteration $[n]$ ensures that for each $[n]$ the overall Abelian constraint, 
as well as the associativity and commutativity of $\odot$ are maintained at each iteration $[n]$ throughout the deformation flow.
 Had one instead expanded the $X_a$ in a power series in $\lambda$ as  $X_{a}[[\lambda]]$,
 and then tried to calculate the expansion of $\mathcal{F}= \exp\{-(\lambda/2) \theta^{ab} X_{a} [[\lambda]] \otimes X_{b} [[\lambda]] \}$ order by order in $\lambda$ by brute force, 
 one would find it difficult to preserve the associativity of $\odot$ during the process.  In the self-consistent twist procedure $\lambda$ is more than merely a formal expansion parameter,  and, as discussed in the next section,  acquires the status of a dimensionful coupling constant. \\

To present the  recursive solutions of the Seiberg-Witten mapping differential equation in greater detail requires a bit more notation.
 Any fixed $\mathcal{F}$
 defines the $r$-th power in $\lambda$ contribution to $f \odot g$ without concern for gauges by
 \begin{align}
 f(x)\odot^{r} g(x) & \doteq \frac{1}{r!}\left(\frac{\lambda}{2} \right)^{r} \theta^{\mu_{1}\nu_{1}} \cdots  \theta^{\mu_{r}\nu_{r}} (\mathscr{L}_{\mu_{1}} \cdots \mathscr{L}_{\mu_{r}} f(x))
 (\mathscr{L}_{\nu_{1}} \cdots \mathscr{L}_{\nu_{r}} g(x))  \\
 \theta^{\mu\nu} & \doteq \theta^{ab}X_{a}^{\mu} X_{b}^{\nu},
 \end{align}
where the standard (non gauge covariant) Lie derivative is employed.
We will use superscripts unenclosed by parentheses or
 brackets to denote the (non-summed) fixed power of $\lambda$ for a fixed twist at which a quantity has been calculated. \\
 
 As previously demonstrated \cite{Ulker}, the Seiberg-Witten mapping leads to the all-order (in $\lambda$) for fixed twist 
 (fixed $[n]$ and fixed $\theta^{\mu\nu}$ field) recursive solution for matter fields
 (suppressing the GM index $a$ on $\phi_a$ and on the gauge potential, as well as the twist iteration superscript $[n]$ for clarity), and now taking $n$ to denote the power of $\lambda$:
 \begin{align}
\hat \phi^{n+1} & = -\frac {\lambda^{n+1}}{4(n+1)} \theta^{\kappa\lambda} \sum_{p+q+r=n} \hat{A}^{p}_{k} \odot^{r}(\partial_{\lambda}\hat{\phi}^{q} + (\hat{D}_{\lambda}\hat{\phi})^{q}),\;\;\;
\mathrm{with} \\
 (\hat{D}_{\mu}\phi)^{n} & \doteq \partial_{\mu}\hat{\phi}^{n} -i \sum_{p+q+r=n} \hat{A}^{p}_{\mu} \odot^{r}\hat{\phi}^{q}.
 \end{align}
 Here $\phi$ is assumed to transform under the fundamental representation of the gauge group, 
but similar results obtain under the adjoint representation.The implied all $\lambda$ order summed deformed scalar field for fixed twist is
 \begin{align}
 \hat{\phi}^{(n+1)} \doteq &\; \phi^{0} + \hat{\phi}^{1} + \cdots + \hat{\phi}^{n+1} \\
  = & \; \phi^{0} -(1/4)\sum_{k=1}^{n+1} (1/k!) ( \theta^{\mu_{1}\nu_{1}}\cdots\theta^{\mu_{k}\nu_{k}}) \Big[\frac{\partial^{k-1}} {\partial\theta^{\mu_{2}\nu_{2}}\cdots\partial\theta^{\mu_{k}\nu_{k} } } 
 \times\nonumber\\
 &\;\; \hat{A}^{(k)}_{\mu_{1}} \odot (\partial_{\nu_{1}} \hat{\phi}^{(k)} + (\hat{D}_{\nu_{1}}\hat{\phi})^{(k)}) \Big]_{\theta = 0}, \\
 (\hat{D}_{\mu} \hat{\phi})^{(n)}  \doteq &\; \partial_{\mu} \hat{\phi}^{(n)} -i\hat{A}_{\mu}^{(n)} \odot \hat{\phi}^{(n)}.
 \end{align}
 Here we have used the same convention from \cite{Ulker}, and taken gauge covariant derivative operators as
 $\partial_{\mu} -iA_{\mu}$. \\
 
 There is a similar set of recursive solutions for the deformed (non-GM) gauge fields $\hat{A}$ in terms of the undeformed $A$. 
 The Seiberg-Witten differential equation for $\hat{A}_{\mu}$ reads ($\{\cdot , \cdot\}_{\odot}$ is the anti-commutator with respect to $\odot$):
 \begin{equation}
 \delta\theta^{\mu\nu} \frac{\partial\hat{A}_{\gamma}}{\partial\theta^{\mu\nu}}   = -(1/4) \theta^{\kappa\lambda} \{\hat{A}_{\kappa}, (\mathscr{L}_{\lambda}\hat{A}_{\gamma})+
 \hat{F}_{\lambda\gamma}\}_{\odot}.
 \end{equation}
 Which for $\theta$ symmetric in its indices reduces to
 \begin{equation}
 \frac{\partial\hat{A}_{\gamma} }{\partial\theta^{\kappa\lambda} } = -(1/8) \{\hat{A}_{\kappa}, (\mathscr{L}_{\lambda}\hat{A}_{\gamma}) + \hat{F}_{\lambda\gamma} \}_{\odot} 
 -(1/8) \{\hat{A}_{\lambda}, (\mathscr{L}_{\kappa}\hat{A}_{\gamma}) + \hat{F}_{\kappa\gamma} \}_{\odot}.
 \end{equation}
 Defining 
 \begin{equation}
 \hat{A}^{(n)}_{\mu} = A_{\mu} + \hat{A} ^{1}_{\mu} +\cdots + \hat{A}^{n}_{\mu},
 \end{equation}
 one has the recursive solution
 \begin{equation}
 \hat{A}^{n+1}_{\gamma} = -\frac{\lambda^{n+1}}{2(n+1)} \theta^{\mu\nu} \sum_{p+q+r=n} \hat{A}^{p}_{\mu}\odot ^{r} \left( (\mathscr{L}_{\nu}\hat{A}^{q}_{\gamma}) + \hat{F}^{q}_{\nu\gamma}\right).
 \end{equation}
 These expressions allow one to obtain the deformed $\lambda$ power summed $\hat{A}$ and $\hat{\phi}$ to arbitrary order in $\lambda$ for any fixed twist.\\
 
 For the remainder of this article we will detour this interesting but computationally intricate deformation flow, 
 and assume that the appropriate self-consistently deformed fields exist and are well defined 
 so we can focus on the physical implications of the model. This is similar to ignoring back-action effects in general relativity.
 
\section{Estimates for nonlocality lengths and particle zoo tour}

Now we turn to study the argument of the exponential in the Abelian twist $\mathcal{F}$, namely $(\lambda/2) \theta^{ab} X_{a}\otimes X_{b}$, by dimensional analysis.
We utilize the fact that the classical (canonical) mass dimension of a quantum field $f$ is $D_{M}(f) = 1+s_f$, where for a field of Lorentz type $(A,B)$, $s_{f}= A+B$. However if
$f$ describes a vector field, and there is a conserved (Noether) current not depending on $f$, then $s_f=0$;  notable examples being the photon and graviton.    
The undeformed interactions we consider in the $XD\phi$ and the $X\gamma\psi$ models are all renormalizable.  Two alternative, physically motivated models will be briefly considered below.
However, they turn out to be non-renormalizable when undeformed, and will be discarded. 
This way we can be assured that
at the energy scale of the $X_a$ (or the fields that comprise them), any ultraviolet divergences can be renormalized or regulated so
there are no energy cut-offs lurking about to upset the dimensional analysis done here. 
Within the classical twist, the $X_a$ are \emph{classical} fields, which should not receive any quantum corrections from fluctuations
if they are computed from stationary points of the matter actions. \\

Let us start with a $XD\phi$ model, that admits $N=1$ or $2$ scalar fields, obeying the $\phi^4$ action, possibly with GM gauge fields. Because we are using dimensional
analysis, the mass models that take $\mathscr{Q}_l = m_l$ have to be treated separately, so we perform the estimate for the other models where $\mathscr{Q}_l$ is dimensionless  first .
Since $D_M(\phi)=1$, so
from equation (\ref{XDPhi}) one has $D_{M}(X_a)=3$.  Alternatively, in the $X\gamma\psi$ model based on Dirac fermion fields $\psi$ of canonical mass dimension $3/2$, one
finds the same result $D_{M}(X_a)=3$  from equation (\ref{XFC}).
Take $\lambda$ to be a real valued dimensionful coupling constant, similar to the Newton-Cavendish $G$ constant or the cosmological constant
 $\Lambda$ entering the Einstein-Hilbert action. 
For the exponent of $\mathcal{F}$ to be dimensionless requires 
\begin{equation}
(M_{\lambda})^{D_M(\lambda)} (M_{\phi})^6 (\xi_c)^{-2} \sim 1, \label{XDPhi_DA}
\end{equation}
where $M_{\lambda}$ is an unknown mass characterizing the couplng constant $\lambda$.  $\xi_c$ is a length scale arising from the Lie derivatives introduced by the $X_{a}$, 
describing the proper distance over which $\mathcal{F}$ introduces nonlocality into the $\odot$-product.  (\ref{XDPhi_DA}) implies $D_M(\lambda)=-8$ for both these models, 
and $\lambda$ then acquires a negative mass dimension, similar to $G$. The resulting order of magnitude estimate for $\xi_c$ is (making no distinction here between $\phi$ and $\psi$)
\begin{equation}
\frac{\xi_c} {L_P} \sim (151.)\, (M_{\lambda}/3\times 10^{3}\,\mathrm{TeV})^{-4} \,(M_{\phi}/\mathrm{TeV})^{3} \mathscr{Q}, \label{XDPhi_xi_est}
\end{equation}
with $L_{P}\doteq (\hbar G/c^3)^{1/2} \simeq 1.6\times 10^{-35}\,\mathrm{m}$ being the Planck length, and  The $X_a$ fields have been approximated by the third power of $M_{\phi}$,
and similarly for $\lambda$. 
The mass models similarly yield
\begin{equation}
\left( \frac{\xi_c} {L_P}\right)_{\mathrm{mass}} \sim (151.)\, (M_{\lambda}/6\times 10^{2}\,\mathrm{TeV})^{-5} \,(M_{\phi}/\mathrm{TeV})^{4}, \label{Mass_xi_est}
\end{equation}
These estimates demonstrate that there is a window of plausible values for $M_{\lambda}$
and $M_{\phi}$ for which the $XD\phi$ and $X\gamma\psi$ models can lead to nonlocality scales $\xi_{c}$ in excess of $L_P$, which is required for classical self-consistency:
 $M_{\phi}, M_{\lambda}\ll (\xi_{c})^{-1}
\ll (L_P)^{-1}$ (in $\hbar =c=1$ units). This general conclusion is robust with respect to nonvanishing anomalous dimensions \cite{Weinberg_AD} entering this analysis, 
as they may alter the exponents in 
(\ref{XDPhi_xi_est}), and thus modify the $(M_{\lambda}, M_{\phi})$ window, but such a window will still exist. Moreover, as previously discussed, 
such anomalous dimensions are expected to vanish
for both the undeformed $XD\phi$ and $X\gamma\psi$ models. This kind of classical $\xi_c$ is precisely what the macroscopic Lieb-Robinson
approach needs to produce microcausality on longer scales.\\

What about the causality issues \cite{Schenkel}\cite{Soloviev}\cite{Szabo} that have long plagued noncommutative geometry, have they been truly banished?
This question can be addressed by examining the scalar field Green's
operators $P$ as studied by \cite{Schenkel}. One finds that for commutative $\odot$-products, the equation for $P$ reduces to the standard Klein-Gordon equation incorporating the
Laplace-Beltrami operator in the presence of the deformed metric. Standard light cone physics remains intact, and scalar fields do not perceive the nonlocality scale $\xi_c$ for mass scales
$M$ such that $\xi_{c} \ll \hbar/Mc$, where the Compton wavelength on the right hand side is the length scale for quantum blurring of the light cone. The banishment appears quite final. \\

These numerical estimates may be used to rule out some of the ways of constructing the $X_a$ from matter fields discussed previously.
Suppose the $X_a$ were composed entirely from SM fields, either by taking $\mathscr{Q}_l$ to be either the particle's (dimensionless) standard $U(1)$ electric charge or unity.
The twist generated by the electron field has to have $\xi_c/L_{P} > 1$ to possess a classical nonlocality length. This implies $M_{\lambda} < 200\; \mathrm{GeV}$.
But then the twist produced by the mass 174 GeV top quark would yield a nonlocality scale $\xi_c$ longer than $4.3\times 10 ^{-19}\;\mathrm{m}$ or $(460\; \mathrm{GeV})^{-1}$
 at its Compton radius $\Lambda_C$,
making it nonlocal and problematic for it to be causally behaved at experimentally accessible energies 
since one no longer has $\xi_{c} \ll \Lambda_C$. Next consider a SM mass model with $\mathscr{Q}_{l}=m_l$.  
A similar estimate using equation (\ref{Mass_xi_est}) for the electron to first constrain $M_{\lambda}$ by $\xi_c/L_{P} > 1$ then yields 
 $\xi_c$ for the Higgs of order $1.3\times 10^{-13}$ m, also long enough to make the Higgs start to act nonlocally in experiments. 
SM mass models with $\mathscr{Q}_{l}=(m_l)^n$, with fixed power $n>1$, only make this problem worse.
Likewise, considering a fermionic model coupling only to electronic lepton number, and using a mass of 0.32 eV for 
the electron neutrino to bound $M_{\lambda}$, leads to an experimentally unacceptable $\xi_c$ in excess of $1.3\times 10^{-15}$ m 
for the electron. However, a SM
fermionic model coupling to baryon number $B$ survives a similar numerical trial when applied to the proton and top quark.
Consequently, either all the SM baryons produce a twist or none, and the mesons generate none.\\

It is also possible to use symmetries to further restrict the possible $\mathscr{Q}_{l}$ entering the $X_a$. 
In particular, let us examine the weak interaction and its $SU(2)_L$ invariance. 
One finds that $\mathscr{Q}_{l}$ must be none of: the electron number(s), the neutrino
number(s), the proton number, or the neutron number; otherwise the deformed action would be $SU(2)_L$ variant, and 
so would the deformed Einstein equation. This happens since the (left-handed parts of) the u and d quarks, 
as well as the electron and its neutrino,
 gauge transform as $SU(2)_L$ doublets. Similarly $\mathscr{Q}_{l} = (B-L)_l$ is ruled out by the numerical analysis for the 
electronic lepton number just discussed. However $\mathscr{Q}_{l}$ could still be the sterile (right handed) neutrino number or the baryon number 
$B_l$ for species $l$ (within the SM).  \\

Non-standard GM  matter could have (non-$U(1)_{SM}$)
$U(1)_{GM}$ interactions, either with or without additional non-Abelian gauges. 
In particular, unbroken $U(1)$ gauge theories in hidden or dark
sectors have been previously investigated by several researchers.
\cite{Gubser_Peebles}\cite{Feng_Kumar}\cite{Feng_2}\cite{Dobrescu} \cite{Pospelov_etal}\cite{Ahluwalia_Lee__etal}\cite{ABCK}
Interestingly, such gauge theories are \emph{naturally admissible} within commutatively deformed general relativity produced by the GM matter sector,
from whose twist nonlocality the microcausality of classical spacetime emerges by the Lieb-Robinson route.   A GM sector, however, is not required to be $U(1)_{GM}$
interacting, and it could still engender a twist via a numerical or mass model that does need $U(1)$ charge. \\

We will now take the reader on a brief guided tour of the GM particle zoo for the separate 
cases of the GM scalar and Dirac fields, which might include spontaneous breaking
of ground state global symmetries.\cite{Kibble} 
Some of these particles could be dark matter (DM) candidates, and
a more detailed analysis of the GM particles' viability as DM candidates will be presented later.
The matter content of the GM sector is the same for $N=2$ as for $N=1$, the only difference being that for $N=2$ one has to build $X_T$ from $X_M$
and the Abelian constraint by the Cauchy-Kovaleskaya construction discussed earlier.
\\

\textbf{GM Scalars: } \\

Turning first to the scalar $XD\phi$ model: We note that new scalars will not upset the SM's delicate gauge anomaly cancellations, regardless of whether or not they carry non-singlet
SM gauge representations. At the same time, the SM does not furnish any stable scalars, and that stability is a requirement if a 
current constructed from that field is to generate the twist in the gravitational sector. So such GM scalars are required to be non-SM fields.\\

The $\phi$ are self-interacting via gravitation, the $\phi^4$ term in the Lagrangian, and also any possible gauge fields $A_{\mu}$ entering the 
currents produced by $\phi$ in the construction given by (\ref{XDPhi}), or its $q_{l}\mapsto \mathscr{Q}_{l}$ extensions.   
These gauge interactions could include any of the known SM gauge fields,
and/or ones coupling only to the GM sector, and will not violate $Z_2$ symmetry in the model
described by (\ref{Phi_4}). For instance a non-SM ``dark" $U(1)_{GM}$ interacting only with these GM scalars by a 
version of scalar electromagnetism would be admissible.
The $\phi^4$ interaction can alter the number of $\phi$ particles if $\phi$ is  a self-adjoint scalar.
If there are additional $Z_2$ symmetric, non-gauge, direct couplings to the SM sector (not included in equation (\ref{Phi_4})), such as $\phi\odot
\phi\odot H^{\dagger}\odot H$, where $H$ is the SM scalar Higgs weak $SU(2)$ doublet,
that symmetry will give isolated single $\phi$ particles protection against decays into purely SM products. 
If $m^2 <0$, the $Z_2$ symmetry is spontaneously broken, and single $\phi$ particles could then decay into suitably coupled, energetically allowed SM products. 
Depending on $\phi$'s gauge couplings, 
there could be a Higgs mechanism in this tachyonic case. A broken symmetry ground state for $\phi$ will then affect the
expression for the twist generator $X$ via equation (\ref{XDPhi}), since the vacuum expectation $\langle\phi\rangle$ acquires a non-vanishing,
spacetime position independent value.  \\

\textbf{GM Fermions:}\\

Next we consider the fermionic $X\gamma\psi$ model.
There is no renormalizable four fermion interaction, but it is expected to be gravitationally self-interacting 
(so it would clump astrophysically), and it could also interact via $U(1)_{GM}$ and/or non Abelian gauge fields. 
 If the GM fermion carries no SM gauge degrees of freedom,
the only renormalizable direct (non-gauge) interaction with the SM sector would have the form $\bar{\psi} \psi H$, with $H$ being the Higgs scalar. However that would violate SM $SU(2)_L$ symmetry. If $\psi$ were a non-SM left handed fermion, and it also carried nonzero SM weak hypercharge $y$, then there would be a gravitational anomaly 
in the SM.\cite{Weinberg_Grav_anomaly} So this (these) fermion(s) is (are) at least one of:
  \emph{all} the SM baryons coupling via $\mathscr{Q}_{l} = B_l$, right handed, or carry no weak
 $U(1)$ hypercharge $y$. It must also be stable (or possess a stable member within its family) in order
 to be of relevance in the present epoch as a twist generator. 
Considering first a right handed fermion such as a light sterile neutrino, this is generally considered to lie in the mass range 1 keV
to about 10 MeV.\cite{Abazajian_01} To keep $\xi_c/L_{P} \sim 100$ then would require $M_{\lambda} \sim (0.056 - 560) $ GeV. 
A heavy sterile neutrino could also have mass in excess of 45 GeV, necessary to maintain the theoretical-experimental agreement for the $Z^0$ total decay rate. Such a heavy neutrino would 
have $\Omega_S h^{2} \lesssim 10^{-3} << \Omega_{DM} h^{2} \simeq .106\pm 0.08$, so it could not constitute most of DM, but could still generate the twist.\cite{Weinberg_Cosmol_P190}
The stable SM possibility for a twist producing fermion is the baryons in stable configurations, such as the proton, 
coupling to baryon number $\mathscr{Q}_{l}=B_{l}$.
Although any stable spin 1/2 nuclei are also conceivable as candidates for this twist generator, 
they cannot be the only source for the twist since
they were not present before Big Bang nucleosynthesis or even later.  But it nevertheless remains a possibility that the familiar SM baryons could 
be the twist generating fermions hiding in plain sight.
We will examine the case that this fermion is a non-SM $y=0$ particle in more detail below.\\

Because of the Fermi statistics, the case $m^2<0$ for fermions does not lead to condensates unless there is some kind of pairing as in BCS superconductivity, superfluid $^3$He or the QCD quark condensate. Also since there is no renormalizable
four fermion interaction to take the role of the scalar $\phi^4$ interaction, this case becomes unstable (its Hamiltonian is unbounded below), so the tachyonic fermions are not physically interesting. \\

Because one is now possibly introducing new fermion fields, one must be cautious about upsetting the delicate gauge anomaly cancellations of 
the SM.\cite{Weinberg_Grav_anomaly} Specifically, this means
\begin{equation}
D_{\alpha\beta\gamma}\doteq \frac{1}{2} \mathrm{tr} \,\big(\{T_{\alpha},T_{\beta}\}\, T_{\gamma} \big) =0,
\end{equation}
where $T_{\alpha}$ is the representation of the gauge algebra on all left-handed fermion and anti-fermion fields, and $\mathrm{tr}$ is a sum over those species.
Additionally, there are possible gravitational anomalies (violations of diff covariance), 
whose absence requires $\mathrm{tr}\,\, T_{\alpha} =0$ for all $U(1)$ gauge generators. 
The conclusions of such an analysis are the following: Recall the SM weak hypercharge defined as $y/g'\doteq t_{3}/g -q/e$, 
where $t_{j}$ for $j=1,2,3$ are the generators of weak isospin $SU(2)$ with coupling constant $g$, $q$ is the particle's electric charge, 
and $g'$ is coupling constant to the electroweak $U(1)$ generator $y$.
If the SM weak hypercharges $y$ of the GM fermion(s) all vanish, then there are no  gravitational $U(1)_{SM}$ gauge anomalies in the SM. 
If there is a (non-SM) $U(1)_{GM}$ gauge interaction in a fermionic GM sector, and if any $U(1)_{GM}$ hypercharge $y'$ vanishes for all GM and SM fermions,
then there are no gravitational $U(1)_{GM}$ gauge anomalies. It is also easy to show from the absence of a $SU(2)-SU(2)-U(1)_{GM}$ gauge anomaly
that either both (electron, neutrino) and (u,d) quarks or neither $SU(2)$ doublet have $y'\ne 0$, consistent with the absence of gravitational gauge anomalies.
It is therefore possible to take $y=0$ for all fermionic GM particles and $y'=0$ for both the GM and SM sectors, 
which will be the starting point for the viability of this case as DM and for comparisons to observations in a 
later section. Setting $y$ to be zero for GM fermions also avoids the complication of mixing between any 
$U(1)_{GM}$ and $U(1)_{SM}$ gauge fields; that is there will be no term in the Lagrangian density of
 the form $(F_{GM})_{\mu\nu}\,(F_{SM})^{\mu\nu}$.\cite{ABCK} \\

For completeness we next examine two alternative models for how to construct self-consistent twist generators $X_a$ from nonscalar matter fields.  
One possibility is simply to take the vector fields $X_a$ to be massive (nongauge) vector bosons, having canonical mass dimension $2$. These bosons
might be described by an action of the form
\begin{align}
S[X] & = \sum_{a=1}^{N} \int_{\mathcal{M}} \mathrm{d}^{4}x \;|g|^{1/2} \odot \left( -\frac{\gamma_{a}}{4} F^{a}_{\mu\nu} \odot  F_{a}^{\mu\nu} - \frac{m_{a}^{2}} {2}
X^{\mu}_{a} \odot X^{a}_{\mu} \right) \\
F^{a}_{\mu\nu}  & \doteq \nabla_{\mu} X^{a}_{\nu} - \nabla_{\nu} X^{a}_{\mu}.
\end{align}
This yields  estimate for the nonlocality
scale $\xi_c$:
\begin{equation}
\xi_{c}(\mathrm{vector\; boson})/L_{P} = (4.6\times 10^5) (M_{X}/\mathrm{TeV})^{2}
(M_{\lambda}/3\times10^{3}\,\mathrm{TeV})^{-3}.
\end{equation}
However, while the second term has has a coupling that superficially resembles a mass, $m_{a}$ actually has mass dimension $0$. The Yang-Mills kinetic term has a coupling constant $\gamma_{a}$
with mass dimension $-2$, making this a nonrenormalizable field theory on the basis of power counting. We therefore discard it.  \\

As a final possibility, consider a Universe with nonzero cosmological constant $\Lambda$. The twist exponent might be conjectured to be
$(\lambda/2) \Lambda g^{\mu\nu} \partial_{\mu} \otimes \partial_{\nu}$. However $g^{\mu\nu}$ generally has too many degrees of freedom to be written in the form
$\theta^{ab} X^{\mu}_{a}X^{\nu}_{b}$ using $N\le 2$ Abelian vector fields $X_a$. So this model would have a non-Abelian twist, resting on shaky mathematical foundations with a generally nonassociative product.
One way out is if the Universe were highly symmetrical, and the $X_a$ could then be chosen to be the Killing vectors of the appropriate de Sitter (or anti-de Sitter)
 Friedmann-Lema\^itre-Robertson-Walker geometry.
\cite{Schenkel}
Then using the $\Lambda$CDM (cold dark matter) model value $M_{\Lambda} \simeq 8.94$ meV would imply the estimate
\begin{equation}
\left(\frac{\xi_{c}}{L_P} \right)_{\mathrm{\Lambda}} \sim (890.) (M_{\Lambda}/8.94\,\mathrm{meV})(M_{\lambda}/\mathrm{TeV})^{-2}.
\end{equation}
So generating the twist from the cosmological constant  could nevertheless produce a plausibly valued $\xi_c$ for a range of $M_{\lambda}$, but only in highly symmetrical cases. Its renormalizablity is also an issue, since the metric tensor generating the twist comes from the 
Einstein-Hilbert action, which is well known to be nonrenormalizable. Due to these deficiencies, we also abandon this way of constructing the twist. \\

As an alternative to coupling SM to GM matter by having GM matter carry nontrivial SM gauge group representation indices, one could imagine there might be direct couplings between the scalar GM and SM sectors.\cite{ABCK}\cite{Cirelli}
One could, for example, couple the SM Higgs $SU(2)$ doublet $H$ to the scalar GM sector's $\phi_a$ by including a (renormalizable) term in the Lagrangian density of the form
\begin{equation}
\mathcal{L}_{\mathrm{SM-GM}} = g_{\mathrm{SM-GM}} \sum_{a=1}^{N} (H^{\dagger} \odot H)\odot (\phi_{a}\odot\phi_{a}),
\end{equation}
with a dimensionless coupling constant. This would still preserve the GM sector's $Z_2$ discrete and $O(N)$ gauge symmetries, as well as SM $SU(2)$ invariance.
Simple GM scalar $\phi$ to SM fermion $\psi_{SM}$ couplings are also restricted: $\bar\psi_{SM}\, \psi'_{SM}\, \phi$ violates $Z_2$ symmetry 
and GM charge conservation; 
and $\bar\psi_{SM}\, \psi'_{SM}\, \phi \,\phi$ is nonrenormalizable. 
Similarly, the familiar Yukawa couplings of two SM $SU(2)$ doublets to a single GM sector fermion would violate both $U(1)_{SM}$ gauge invariance 
and any possible GM charge conservation.  Notice that the amplitude of any process driven
by these kinds of terms will be (to zeroth order in the twist) independent of twist parameters such as $\theta^{ab}$ or $M_{\lambda}$. 
We will discuss SM-GM couplings further below when the astrophysical implications of GM matter are discussed. \\

Just how small are the deformations we have been considering? For instance, first consider atomic or nuclear spectroscopy. A simple order of magnitude estimate 
for the twist induced relative first order change $\Delta H$ in the energy level spacings $H_{0}$ is
\begin{equation}
| \Delta H/H_{0} | \sim (\xi_c/L_{\mathrm{char}})^2,
\end{equation}
where   
$L_{\mathrm{char}}$ is some characteristic size of the process. Using $\xi_{c}/L_{P} \sim
10^{2} - 10^{5}$, one finds that atomic electronic transitions have relative twist induced changes of order $10^{-46} - 10^{-40}$. Nuclear or hyperfine transitions display relative changes
of $10^{-36} - 10^{30}$. Particle physics branching ratios in the TeV range have relative perturbations of order $10^{-28} -10^{-22}$. Such miniscule changes are significantly beyond present experimental detectability. One way to understand this is that the nonlocal effects
are practically of Planck scale. \\ 

Using the above models one may also estimate the relative size of change a single twist produces on the $X$ fields themselves (twist self-dependency) 
to first order in $\lambda$ as
\begin{equation} 
\eta \doteq \frac{X^{1}} {X^{0}} \sim (10^{-28}) (M_{\phi}/\mathrm{TeV})^{8} (M_{\lambda}/3\times10^{3}\,\mathrm{TeV})^{-8}.\label{FracX}
\end{equation}
The presence of the twist means the variation of the action $\delta S/\delta\psi$ receives contributions from $\delta \odot/\delta\psi$, which are deformed sources for any field $\psi$, 
even on shell.
Such sources have relative size smaller than the standard ones by at least one power of
 $\lambda$. In the gravitational sector, the twist has even greater numerical efficiency:
A simple order of magnitude calculation utilizing either the scalar or fermion model yields the first order deformed metric tensor as
\begin{equation}
| g^{1}_{\mu\nu} | \sim \left(\frac{\Phi_{\mathrm{grav}}} {M_{\lambda} L_{c}} \right) \left( \frac{M_{\phi}} {M_{\lambda}} \right) ^{7}.
\end{equation}
Here $\Phi_{\mathrm{grav}}$ is the dimensionless Newtonian gravitational potential, and $L_c$ is the (minimum) radius of curvature of  spacetime at the event in question.
Using $M_{\lambda} = 3\times 10^{3}$ TeV and $M_{\phi}=1$ TeV, the first order deformation
part of the metric tensor at the Earth's surface is $|g^{1}| \sim  10^{-67}$. 
This is about $10^{58}$ times smaller than the Newtonian contribution to the metric tensor at the reader's present location.
Traveling at the speed of thought to another astrophysical extreme,
at the event horizon of a $10$ km radius black hole one finds $|g^{1}| \sim 10^{-51}$. Clearly geodesics will be \emph{very} subtly perturbed from their undeformed courses.
These tiny deformations lend some credence to the assumption that deformation flow will converge to a self-consistent twist solution for astrophysical situations. \\

Next we turn to renormalizability of the twisted field theory. The role of the twist is to introduce arbitrary many powers of $\Delta L \doteq
-(\lambda/2) \theta^{ab}
X_{a}^{\mu} X_{b}^{\nu} \hat{\mathscr{L}}_{\mu} \otimes \hat{\mathscr{L}}_{\nu}$ into the undeformed Lagrangian. Here the $X_a$ are classicized fields, and $\lambda$ has canonical mass dimension $-8$. Even though these insertions by themselves are 
of total mass dimension zero by construction, their Lie derivatives $\hat{\mathscr{L}}_{\mu}$ act on the other fields in the Lagrangian to generate terms in the deformed Lagrangian with sufficient derivations to become nonrenormalizable. That is, those terms
will have an overall coupling constant of negative mass dimension. Hence the deformed Lagrangian
is a nonrenormalizable effective quantum field theory. However, renormalizability is \emph{not}
a fundamental requirement for a physical theory. As lucidly discussed by S. Weinberg in ``Is
Renormalizability Necessary?"\cite{Weinberg_Ren_Nec}, as long as one includes in the Lagrangian
all of the infinite number of interactions allowed by the symmetries, then there will be counterterms available to cancel every UV divergence. The twist will do this because of the systematic way it inserts arbitrarily many factors of $\Delta L$ into the undeformed action.
Then on dimensional grounds, the terms having couplings with negative 
mass dimension $g_{i} \simeq M_{i} ^{\Delta_{i}}$, with $\Delta _{i}<0$ and $M_{i}$ some mass characterizing the $i$-th interaction, will have their effects suppressed for momenta 
$k \ll M_i$ by a factor $(k/M_{i})^{-\Delta_{i}} \ll 1$. Einstein-Hilbert gravity is well known to be nonrenormalizable, with an characteristic energy scale of order the Planck energy $E_P$. So what is the effective energy scale of the deformations being considered here? 
The deformation's single insertions are of the form $\xi_c^{2}
 (\mathscr{L} F_{1}) \cdot (\mathscr{L} F_{2})$, where $F_{1}, F_{2}$ are factors of the Lagrangian outside this twist's single action. Therefore the energy scale of the deformations are on the scale $\xi_{c}^{-1} \sim (10^{-5} - 10^{-2}) E_{P}$, that is close to, but not at, the Planck scale $E_P \sim 8\times 10^{16}$ TeV. Notice this is \emph{not} the other scales entering the twist: $M_{\lambda}\sim 10^3$ TeV or $M_{\phi}\sim $ TeV. The nonrenormalizable effects at momentum $k << \xi_{c}^{-1} $ are then suppressed by a factor $(k\xi_{c})^{2} \simeq (10^{-28} - 10^{-22}) (k/\mathrm{TeV})^{2}$. As $k$ approaches $\xi_{c} ^{-1}$ from below, and starts to exceed it, there will be nonunitary contributions to the S-matrix in this model. These arise from the new \emph{quantum} character of the nonlocality at those scales: the $X_a$ can no longer be taken as coming from classical stationary points or expectation values as this model does, and the twist becomes a quantum object, along with spacetime itself. This emergence of new physics at a scale corresponding to that of nonrenormalizability is roughly similar to what occurs at the electroweak scale of about 300 GeV, and is well beyond the scope of this article. \\
 
\section{Dark matter implications of classical commutative deformations}

Commutatively deformed classical gravitation has several interesting potential implications for  astrophysics, cosmology, and quantum gravity. Foremost among these is the possibility that the GM
matter could be DM. To summarize that relationship so far: GM matter acts just like SM matter from a gravitational point of view; it responds to and acts on both GM and SM matter in the classical way. Therefore it will clump gravitationally, provided it is nonrelativistic in a cosmologically comoving frame. Due to their expected TeV range mass, similar to weakly interacting massive particles (WIMPs), that is likely to be the case. Stability over cosmological time scales is required of GM matter to generate the twist and for DM as well. GM matter possesses no a priori strong, electromagnetic, or weak interactions; however nonminimal coupling is not precluded, and in particular it could be weakly interacting.\cite{ABCK}\cite{Cirelli} DM is observationally known to be neither electromagnetically nor strongly interacting. In the following we will examine both minimal (totally non-SM interacting) GM matter and weakly interacting non-minimal GM matter as DM candidates.  Measurements of spin independent elastic scattering cross sections of weakly interacting DM off nuclei have ruled out DM particles with nonzero weak hypercharge $y$ unless appropriate weak couplings are introduced\cite{Cirelli}, consistent with the assumption that GM has $y=0$, as discussed earlier. Additionally, several of the GM models are self interacting, and observations show DM to have some self interactions $\sigma/M \lesssim 0.1 - 1.25$ cm$^{2}$/g, while maintaining their collisionless galactic dynamics.\cite{ABCK} To be viable as a DM candidate, the GM must also pass a well-known abundance test \cite{Bertone}\cite{Dodelson}, which will be performed during the following pass through the GM sector from the DM perspective. \\

DM abundance estimates assume that at some time in the early Universe DM was in local thermodynamic equilibrium both with itself and with the other constituents. Later on the reaction
rates for processes maintaining these equilibria fell below the Hubble expansion rate, and then those processes became ``frozen out." The time scale for the reaction grew longer than the age of the Universe. If a reaction had differing numbers of DM matter particles
on its left and right sides, then it can contribute to the measured abundance of DM. Reactions that have the same number of DM particles on both sides
do not affect the abundance of DM, although they can play a role in DM self interactions. 
The analysis of the abundance determining reactions is handled by a Boltzmann equation approach, where the 
quantum statistics of the particles is ignored because the densities are far from the quantum degeneracy regime. The result is a relationship expressing the DM abundance in terms of the particle mass, its thermally averaged cross section $\sigma_{0}\doteq\langle\sigma v\rangle$, and the number of degrees of freedom at freeze out. Here $v$ is the relative velocity of the annihilating particles in a cosmologically comoving frame in $c=1$ units. A given DM model, together with the measured DM abundance $\Omega_{DM} h^{2} \simeq 0.106$ then yields $\sigma_{0}$ in terms of the particle's mass, or vice versa if one can independently calculate the cross section.\cite{ABCK}\cite{Cirelli} This
can be made more complex by resonant enhancement effects (co-annihilations), which are ignored below.\cite{Bertone} We now consider GM sector case by case to identify these processes. \\

As discussed earier, the value of $N$ does not affect the matter content of the GM sector. We first turn to the GM scalars, which are necessarily non-SM particles: \\

Real scalar GM matter without GM charges, GM gauges, or weak interactions has its abundance determined from reactions of the form $\phi\phi\phi\leftrightarrow\phi$ arising from the $\phi^4$ interaction in the Lagrangian. Similarly that term will also generate short range GM-GM interactions, which can freeze out. Single $\phi$ particles can be $Z_2$ stabilized against decay. In order to agree with the observed DM abundance, one finds this case yields $(M/\mathrm{TeV}, \sigma_{0}/\mathrm{pBarn})$ values $(0.1,0.68)$, $(1.0, 0.76)$, $(10., 0.89)$, and $(100., 0.91)$. The picoBarn characteristic size of $\sigma_{0}$ is of the same order as calculated from WIMP DM models. If $\phi$ is not weakly interacting, this could be used to bound the size of the $\phi^{4}$ coupling constant, which is left for future research.
If one allows $\phi$ to interact by SM weak interactions, then it was shown in \cite{Cirelli} that the simplest $y=0$ scalar would be a weak $SU(2)$ triplet, its electrically neutral particle would be the GM $\phi$ having a dark matter abundance determined mass of $2.0$ TeV, coming from its calculated weak cross section. It is accompanied by two oppositely electrically charged particles having a mass 166 MeV larger, which would decay into the lighter neutral particle. This mass is in accord with the above estimate derived from the twist. $Z_2$ symmetry would block 
$\phi\rightarrow H^{\dagger} H$ 
decays, but not $\phi\phi\rightarrow H^{\dagger} H$, which would determine the $\phi$ abundance. The tachyonic case $m^2<0$, would be unstable with respect to the first weak decay channel if 
$M(\phi)\ge 2M(H) \simeq 250$ GeV, and then tachyonic $\phi$ would not be a viable DM candidate. \\

As an alternative, a GM charged scalar $\tilde{\phi}$ interacting via a $U(1)_{GM}$ gauge field would be complex valued,
and there will be no $\phi^4$ mediated 
$\phi\phi\phi\leftrightarrow\phi$ reaction since that would violate GM charge conservation (GM gauge invariance). 
However these scalars interact with massless dark $U(1)_{GM}$ photons $\hat{\gamma}$, so pair
annihilation $|\tilde{\phi} |^{2}  \leftrightarrow 2\hat{\gamma}$ can determine the freeze out abundance, 
even in the absence of weak interactions. An analysis similar to the real scalar 
finds $\sigma_{0}$ varies from $0.70$ to $0.93$ pBarn as the GM particle mass
ranges from $0.1$  through $100$ TeV. This could be used to derive the dark scalar fine structure constant (GM charge) as a function
 of $M_{\phi}$, provided the Feynman rules for scalar electrodynamics are in hand, 
a project off the main track of this article and left for later study.
If this scalar couples to the SM weak interaction, then the $Z_2$ allowed process $|\tilde{\phi} |^{2}  \leftrightarrow H^{\dagger} H$
is also a possible abundance determining reaction. $\tilde{\phi}\leftrightarrow H^{\dagger} H$
violates $U(1)_{GM}$ gauge invariance. $Z_2$ symmetry is no longer required to stablize DM in this case since
the lowest mass GM particles carry conserved GM charge. DM becomes a GM plasma, displaying scalar dark $U(1)$ electromagnetism.
 For $m^2<0$, there are no more long range GM gauge interactions since the $U(1)_{GM}$ gauge boson becomes massive,
 and a GM neutral Higgs $\delta\phi$ emerges.
If $\delta\phi$ is nonrelativistic in a cosmologically comoving frame, there is no energetically allowed
abundance determining process without weak interactions. 
With weak interactions, $\delta\phi\rightarrow H^{\dagger} H$ is allowed if $M(\delta\phi)\ge 2M(H) \simeq 250$ GeV since $Z_2$ is broken, 
and  $\delta\phi^{2}  \leftrightarrow H^{\dagger} H$ is also allowed if $M(\delta\phi)\ge M(H) \simeq 125$ GeV. 
The first decay mode could render tachyonic $\tilde{\phi}$ weakly unstable, and then it would be an implausible DM candidate. \\

Now we turn to the fermionic twist models.\\

As discussed earlier, the twist generating fermion could be a sterile neutrino, a light stable SM particle coupling to baryon number, 
or possibly DM in the form of a non-SM particle. Here we examine the last possibility. In the absence of a renormalizable four fermion process and gauges,
there are no abundance determining reactions without coupling to the weak interactions,
hence no implied relationship between averaged cross section and particle mass.
One can estimate the mean free path $\ell$ of these inert $\psi$ particles due 
 to their mutual scattering by $\ell\simeq 1/(n\sigma)$, and crudely estimate
 the cross section's order of magnitude using $\sigma \sim (\hbar/M_{\psi}c)^2$. At the present density of dark matter, this gives $\ell \sim (2.8\times 10^5\,\mathrm{Gly})(M_{\psi}/
 \mathrm{GeV})^{3}$. This estimate renders this GM fermion essentially noninteracting, and it would be problematic for the $\psi$ to thermalize among themselves. Thus without being weakly interacting, this fermion would be an undesirable DM candidate. Hence we look to the weak coupling of $\psi$. Following Cirelli, Fornengo, and Strumia (hereafter CFS) \cite{Cirelli}, the simplest way to accomplish this is to make $\psi$ one member of a SM weak $SU(2)$ triplet, the other two particles being $U(1)_{SM}$ (electrically) charged. Requiring agreement with the observed dark matter abundance, CFS find $M(\psi) = 2.4$ TeV, consistent with the twist estimate (\ref{XDPhi_xi_est}), and lying $166$ MeV below the charged members of the triplet.\\

A $U(1)_{GM}$ GM charged fermion, together with its dark massless $U(1)_{GM}$ photons $\hat\gamma$, will have long range interactions, 
and its abundance determining process without weak interactions is pair annihilation: $\bar{\psi} \psi\leftrightarrow 2\hat{\gamma}$.  
As in the complex scalar case, $Z_2$ symmetry is no longer required to stabilize the GM sector. This dark version of SM spinor 
electromagnetism is an appealing GM model for DM, a GM plasma similar to the complex scalar case just discussed. 
Its phenomenological consequences were investigated by L. Ackerman, M. R. Buckley S. Carroll, and M. Kamionkowski (hereafter ABCK) \cite{ABCK},
motivated by theoretical considerations of unbroken $U(1)$ gauges in 
hidden sectors.\cite{Gubser_Peebles}\cite{Feng_Kumar}\cite{Feng_2}\cite{Dobrescu} \cite{Pospelov_etal}\cite{Ahluwalia_Lee__etal} 
ABCK demonstrated that there are no constraints from either Big Bang nucleosynthesis or the cosmic microwave background
on the additional relativistic degrees of freedom from dark radiation or from nonrelativistic
(heavy) degrees of freedom  in this model. These bounds arise because the Big Bang nucleosynthesis of the observed ratios of cosmic 
nuclear abundances are very sensitive to the expansion rate of the Universe at that time ($T\sim 1$ MeV), which is related to the energy 
density of radiation by the Friedmann equation.\cite{Feng_2} In fact, the analysis of ABCK is readily applied to all the other DM 
models considered here, with the same conclusion.  They further showed that the abundance bounds on the DM annihilation to dark photons 
are inconsistent with collisionless DM in galactic dynamics unless $M_{\psi} > 10^{2}$ TeV, where the dark  spinor electrodynamical fine 
structure constant $\hat{\alpha}$ becomes nonperturbative. That is, the dark photons  interact among themselves very effectively via GM 
matter loop mediated processes.  However, ABCK could achieve the observed DM abundance at lower $\hat{\alpha}$ by opening weak 
annihilation channels. Even in the presence of those weak channels there is no dark photon to visible photon mixing, there is vanishing 
leading order dark photon coupling to SM fermions, and the lowest order of that coupling is proportional to $\alpha^{2}\hat{\alpha}$ coming 
from a GM loop.  \\

To summarize:  Both
GM fermionic models would necessarily have to be SM weakly interacting in order to describe DM. Twist fermions need not be DM, and alternatively
might either be a light stable SM particle coupling to baryon number or a sterile right-handed neutrino. 
More work is necessary to determine whether the $m^2 >0$ scalar GM models also need to be coupled to SM weak interactions. 
The scalar tachyonic models are not likely to be good DM candidates.
When the models interact via the SM weak interaction,
the observed DM abundance combined with the calculated weak cross section gives a particle mass
in the low TeV range where the twist produces a nonlocality scale $\xi_c$ on the order of 
$10^{2} L_P$.
The GM charged twist models provide a theoretical basis for DM as $U(1)_{GM}$ gauge interacting plasmas 
of GM scalars or spinors, which do not require $Z_2$ symmetry to be stable. 
They can arise from twist deformed nonlocal classical diffs, which at the same time produce
microcausality at proper length scales longer than the length  $\xi_c$ characterizing the nonlocality.  
More complicated GM matter models might be constructed by incorporating non-Abelian GM gauges, but
there is little observational motivation or guidance for that step at this point.
\\

\section{First order deformed electromagnetic plane wave propagation through GM matter}

One common experimental probe of quantum gravity theories is to search for dispersion and attenuation of electromagnetic radiation propagating through near vacuum over 
cosmological distances, on the order of giga-lightyears (Gly) or longer.\cite{Alfaro}\cite{Fermi_LAT}
Because classical commutative twists will insert new terms into the Maxwell equations for electromagnetic plane
waves traversing the matter generating the twist, one anticipates similar effects could arise. Here we estimate the size of those effects  
to first order in the twist parameter $\lambda$ together with a simple model for the $\phi$ or $\psi$ field dependent couplings. 
This is essentially first order deformed on shell (classical)
electromagnetism. As will explained, this can also be taken as a test of GM matter as DM.\\

In order to simplify the mathematics, it is necessary to make some assumptions. First we assume a flat spacetime so that curvature effects, $|g|^{1/2}$ factors, and so on may be
dropped. The GM fields generating the twist will be found to couple to the electric $E_j$ and magnetic $B_j$ fields through contraction of indices, like the spatial $j$. 
We assume the GM
matter may be modeled as a homogenous and isotropic gas of particles. Consequently, on average all odd order spatial tensors comprised from the GM
 fields, such as GM gauge fields $A_{\mu}$ or 
$D_{\mu} \phi$, vanish. Even order tensors are taken to decompose into isotropic linear combinations of Kronecker deltas times constants depending on expectation values of the GM fields.
We also neglect the variation of those constants in time and along the electromagnetic radiation's propagation path. More sophisticated models of  GM matter will be left for future research.  \\

In the absence of external electromagnetic currents, the action takes the form
\begin{align}
S_{\mathrm{em}} & = S_{\mathrm{undef}} + S_{\mathrm{deform}} \nonumber \\
& =  -\frac{1}{4} \int \mathrm{d}x^{4} \Big\{ (\partial_{\mu}A_{\nu}-\partial_{\nu}A_{\mu})
 (\partial^{\mu}A^{\nu}-\partial^{\nu}A^{\mu}) - \nonumber \\
 & \;\; \;\;\;    \left( \frac{1}{4}\right) \left(\frac{\theta^{ab}\lambda}{2}\right) \partial_{\lambda}\left( \partial_{\mu}A_{\nu}-  \partial_{\nu}A_{\mu}\right)
  \partial_{\sigma}\left( \partial^{\mu}A^{\nu}-  \partial^{\nu}A^{\mu}\right) \left( X^{\lambda}_{a} X^{\sigma}_{b}\right) \Big\}.
	\end{align}
  From this it is straightforward to derive (overdots designate time derivatives)
  \begin{align}
  \dot{E}_{j} -(\nabla\times B )_{j} & = -(\theta_{\mathrm{curl}})_{j} \doteq  \left(\frac{-\theta^{ab}\lambda}{2}\right) \sum_{i=1}^{6} K_{i}, \nonumber\\
  K_1 & =(X^{\lambda}_{a} X^{\sigma}_{b}) (\partial_{\lambda}\partial_{\sigma})  \big( -\dot{E}_{j} + (\nabla\times B)_{j}\big), \nonumber\\
  K_2 & = \partial_{\lambda} \big( -\dot{E} _{j} + (\nabla\times B)_{j}\big) \big[\partial_{\sigma}(X^{\lambda}_{a}X^{\sigma}_{b}) \big], \nonumber\\
  K_3 & = - (\partial_{\sigma}\partial_{t}) (X^{\lambda}_{a}X^{\sigma}_{b}) (\partial_{\lambda}E_{j} ),  \nonumber\\
  K_4 & = \big[\big( \partial_{\sigma}\partial_{k}\big)(X^{\lambda}_{a}X^{\sigma}_{b}) \big]  \big[ \epsilon_{jkl} \partial_{\lambda}B_{l}\big],  \nonumber\\
  K_5 & = - \big[\partial_{k} (X^{\lambda}_{a}X^{\sigma}_{b}) \big] (\partial_{\sigma}\partial_{\lambda}) (\epsilon_{kjl} B_{l}),  \nonumber \\
  K_6 & = - \big[\partial_{t} (X^{\lambda}_{a}X^{\sigma}_{b}) \big] 
  \big[ (\partial_{\sigma}\partial_{\lambda}) E_{j}\big],
  \end{align}
and
\begin{align}
  \nabla\cdot E & = \theta_{\mathrm{div}} \doteq  \left( \frac{-\theta^{ab}\lambda} {2}\right) \sum_{i=1}^{4} L_{i}, \nonumber\\
  L_1 & =  - (\partial_{\sigma}\partial_{\lambda})(\nabla\cdot E) (X^{\lambda}_{a}X^{\sigma}_{b}), \nonumber\\
  L_2 & = - (\partial_{\lambda}\nabla\cdot E) \big[ \partial_{\sigma}(X^{\lambda}_{a}X^{\sigma}_{b})\big],  \nonumber\\
 L_3  & = - (\partial_{\lambda}E_{j}) \big[ (\partial_{\sigma}\partial_{j}) (X^{\lambda}_{a}X^{\sigma}_{b})\big], \nonumber\\ 
  L_4 & = - (\partial_{\sigma}\partial_{\lambda} E_{j}) \big[ \partial_{j} (X^{\lambda}_{a}X^{\sigma}_{b}) \big], 
  \end{align}
  together with the undeformed Maxwell equations $\nabla\cdot B =0$ and $(\nabla \times E)_{j} + \dot{B}_{j} = 0$.  
  Using the homogenous isotropic gas model for the $\phi$ particles discussed above, these may be reduced to
  \begin{align}
  (\theta_{\mathrm{curl}})_{j} & = F \,\dot{E}_{j} + H\, \nabla^{2} E_{j} + G\,\ddot{E}_{j} \,+\, \mathscr{O}(\lambda ||\theta^{ab}||/2)^{2}, \;\;\;\mathrm{and} \\
 \theta_{\mathrm{div}}& = 0 \,+\, \mathscr{O}(\lambda ||\theta^{ab}||/2)^{2}. 
 \end{align} 
 We have defined 
 \begin{align}
 & \alpha^{ab}  \doteq \lambda \theta^{ab}/2 \\
 & F  \doteq \big( (2/3) \partial_{l}\partial_{t}\langle X_{a}^{l} X_{b}^{0}\rangle + \partial_{t}\partial_{t} \langle X_{a}^{0} X_{b}^{0}\rangle + (1/3) \partial_{l}\partial_{s} \langle
 X_{a}^{l} X_{b}^{s}\rangle \big) \alpha^{ab} \\
& H  \doteq (-1/3) \partial_{t} \langle X_{a}^{n} X_{b}^{n}\rangle \alpha^{ab} \\
& G  \doteq \big( (2/3) \partial_{k}\langle X_{a}^{0}X_{b}^{k}\rangle -\partial_{t}\langle X_{a}^{0}X_{b}^{0}\rangle\big) \alpha^{ab}.
\end{align}
The angle brackets denote  volume averages, discussed more in a moment. These yield the first order deformed wave equations
\begin{align}
\nabla^{2} B_{j} -\ddot{B}_{j} & = -(\nabla\times\theta_{\mathrm{curl}} )_{j}, \\
\nabla^{2} E_{j} -\ddot{E}_{j} & = F\,(\partial_{t})^{2} E_{j} +G\,(\partial_{t})^{3} E_{j} + H\, \nabla^{2}(\partial_{t}) E_{j}.
\end{align}
Substituting the plane wave form $E_{j} \sim \exp [-i\omega t + i \vec{k}\cdot\vec{x} ]$ one finally arrives at
\begin{equation}
\omega ^{2} - k^{2} = -F\omega^{2}+iH\omega k^{2}+iG\omega^{3}.
\end{equation}
This may be further decomposed into
\begin{align}
v_{g}-1 & \doteq \mathrm{d}\omega /\mathrm{d} (\mathrm{Re}\,k) -1  = -F/2 +\mathscr{O} (\lambda^{2}), \;\mathrm{and} \label{Defd_Disp} \\
\gamma  & \doteq \mathrm{Im} \,k  =-(1/2) (G+H) \omega^{2} + \mathscr{O} (\lambda^{2}). \label{Defd_Attn}
\end{align} 
The coefficients $F,G,H$ enter the Euler-Lagrange (Maxwell) equations describing the propagation of $U(1)_{SM}$ radiation through a uniform gas of GM particles.
For now we take the GM particles to be dark matter, and use our results to test that hypothesis.
 Since the average density of dark matter in the present epoch using parameters from the $\Lambda$CDM model is $n_{D}\simeq (1.26\,\mathrm{m}^{-3})
(M_{\phi}/\mathrm{TeV})^{-1}$, the typical $\phi$ particle spacing is 
$n_{D}^{-1/3} \sim (1\,\mathrm{m}) \gg \Lambda_{C} \sim (10^{-18}\,\mathrm
{m})(M_{\phi}/\mathrm{TeV})^{-1}$, where $\Lambda_{C}$ is the Compton wavelength of $\phi$.  So one has a dilute classical gas of  widely spaced particles. 
To enforce uniformity, one has to average
the deformed Maxwell or wave equations over a scale at least as large as the typical interparticle spacing. 
Consequently one has $\langle |X_a| \rangle \sim \eta |X|_{\mathrm{pk}}$, with
 $\eta$  denoting the volume filling factor $\eta\simeq n_{D}\Lambda_{C}^{3}\simeq
(9.6\times 10^{-60})
(M_{\phi}/\mathrm{TeV})^{-4} a^{-3}$, and $|X|_{\mathrm{pk}}$ is the value of the $\phi$ field smeared over its Compton radius. $a$ is the cosmological scale factor, set to unity in the present epoch. This type of volume averaging does not occur 
within the twist itself since $\mathcal{F}$ should not depend on averaging over details of its ``environment,'' unlike the electromagnetic wave propagating through a homogenous gas considered here.\\

For the scalar and fermion models, and roughly approximating $\langle X^{2}\rangle \sim \langle |X|\rangle ^{2}$, one obtains the order of magnitude estimate
\begin{equation}
F \sim (1.5\times 10^{-87}) (M_{\phi}/\mathrm{TeV})^{4} (M_{\lambda}/3\times 10^{3}\,\mathrm{TeV})^{-8} a^{-3}
\end{equation}
as a frequency independent index change produced by the $\phi$ particle gas due to its deformation of the classical electromagnetic field. While such a small 
``dispersion'' would be nearly impossible to measure, it is nevertheless consistent with all experimental measurements to date for the absence of vacuum
 dispersion. Even without the smearing factor $\eta$ the dispersion $F$ would be $\sim 10^{-28}$, still immeasurably small.  \\

The corresponding penetration depth $\Lambda$ of electromagnetic radiation  is estimated to be
\begin{equation}
\Lambda  \sim (1.4\times 10 ^{55}\,\mathrm{Gly})(E_{\mathrm{ph}}/\mathrm{MeV})^{-2}
 (M_{\lambda}/3\times 10^3\,\mathrm{TeV})^{8} (M_{\phi}/\mathrm{TeV})^{-3},
\end{equation}
for photon energy $E_{\mathrm{ph}}$. This implies photons will be absorbed (photon observations will be cut off) after traveling cosmological distances $\Lambda$ when
\begin{equation}
E_{\mathrm{ph}} > E_{\mathrm{co}} \sim (4.\times 10^{5} \,E_{\mathrm{Pl}}) (\Lambda /\mathrm{Gly}) ^{-1/2} (M_{\lambda}/3\times 10^{3}\,\mathrm{TeV})^{4} (M_{\phi}/\mathrm{TeV})^{-3/2} a^{3/2},
\end{equation}
where $E_{\mathrm{Pl}}\simeq 7.75\times10^{16}$ TeV is the Planck energy. It is noteworthy that ignoring the
 filling factor by setting $\eta$ to unity would instead yield a cut off energy of
$E'_{\mathrm{co}}\sim (1.2\times 10^{-2}\,\mathrm{MeV}) (\Lambda/\mathrm{Gly})^{-1/2}$. Thus the volume averaging leading to 
$\eta\simeq n_{D}\,\Lambda_{C}^{3}$ is crucial in allowing 
observations of high energy photons arriving at Earth from cosmological distances with energies well in excess 
of $E'_{\mathrm{co}}$, such as those originating from Gamma Ray Bursters 
having photon energies in the TeV range, as have been detected by Fermi-LAT.\cite{Fermi_LAT}
Since $a\sim 10^{-4}$ at photon decoupling (atomic recombination) with $a^{3/2} \sim 10^{-6}$, photons 
all the way up to Planck scale energies have been able to travel gigalightyear distances without twist induced 
absorption ever since. Before then, during the radiation dominated hot Big Bang, the photon attentuation 
distance was severely limited by plasma effects 
well before photons could travel over cosmological length scales. \\

To summarize this section: The first order $\mathscr{O}(\lambda^{1})$ deformation effects of a commutative twist on classical electromagnetism using the $XD\phi$ and $X\gamma\psi$ models,
in conjunction with approximating the twist generating GM particles as an isotropic and homogeneous gas having a number density equal the present epoch's value
for the average dark matter density,
predicts no cosmological dispersive or absorptive effects detectable with the current technology. 
This result is consistent with measurements of vacuum absorption and dispersion made so far, and it supports the viability of 
the twist-based GM sector as dark matter. Taking twist generating fermionic matter to be the proton, and examining
similar effects on photon propagation over $10$ km through a substance like liquid water with number densities 
$\sim 3\times 10^{22}$ cm$^{-3}$,
implies a cutoff photon energy of about $2\times 10^{7}$ TeV, which is not of physical relevance.
There is expected to be an analogous effect of the twist on gravitational wave propagation, 
which calculation has not yet been performed. However, it too is anticipated to be 
experimentally undetectable, particularly since present day gravitational wave observatories search at low frequencies 
(hundreds of Hz, with some proposals at a few GHz). \\
 
 \section{Self-criticism, future directions, and conclusion}
 
Where are the weak spots? One potential source of undesirable mathematical pathologies could be the use
of the differential form of the twist. Similar to Fourier analysis throughout physics and engineering,
this can lead to troubles if the objects it acts on, or is made of, do not have a sufficiently rapid fall off at
large relative distances. Here those objects are the vector generators
 $X_a$, coming from classicized GM matter currents, along with the other SM fields. 
The typical remedy is to appeal to the Riemann-Lebesgue theorem and to require those fields or currents to be $L^{1}$, or to set
the fields to zero outside some $4$-volume or box. 
The latter approach would be inadvisable here since the points
 where $X_a$ vanishes are defects in the $N=2$ models. But one could still insist that the classical $X_1$ be drawn from some suitable 
Schwartz space where they fall off nicely at spatial infinity, but along the way do not vanish. Essentially this imposes a restriction
on the long distance fall off of the fields. Alternatively, one might use an integral kernel in momentum space as an approach to 
constructing the twist, as has been applied to commutative deformations of flat spacetime\cite{Lizzi}, while maintaining background 
independence by constructing that kernel from matter fields, as the $X_a$ have been here. It would certainly be worthwhile to study 
such an approach.\\

The defects where $X_{1}=0$ arising in the construction of the twin $X_2$ for the $N=2$ cases are simultaneously interesting and threatening. 
The implied breakdown of predictability and classical determinism is not new to gravitational physics,
but the defects remain a concern.  They are precisely the points that are not regular in the sense of 
Aschieri and Castellani \cite{AC_Abel}, discussed in section 2. If $X_2$ also vanished at such locations, 
the Lieb-Robinson mechanism would not be able to produce microcausality there since the twist would become trivial. 
If the set of $X_{1}=0$ defects consists of isolated events or has sufficiently small dimensionality or measure, 
one might be able to assign a nonvanishing value to
the twin $X_2$ at those $X_{1}=0$ locations by smoothness, thereby removing the problem.  
However, this is only a conjecture at this point.  Unlike Cauchy horizons or singularities inside black holes, 
these defects are not cloaked by an event horizon. Space travelers take note.\\

Another place where ignorance could have deleterious consequences is the role of nonzero anomalous dimensions
which enter the $X_a$ when they are classicized as expectation values of  quantum currents. 
While these are not expected to enter, an uninvited guest might still
perturb the dimensional analysis for for nonlocality scale $\xi_c$. However, if we choose to
adjust $M_{\lambda}$ and/or $M_{\phi}$ to keep $\xi_c$ in the range $(10^{2}-10^{5}) L_{P}$, 
this would not appreciably affect the twist's estimated experimental consequences, or the the 
nature of the effective field theory. The \emph{existence} of that window for $\xi_c$ is robust with 
respect to those possible anomalous dimensions.\\

The computationally complex deformation flow to a self-consistent twist is practically unexamined, and needs to be understood in greater detail. 
\\

There are several different future theoretical directions to explore.  The scalar 
model needs more work to determine
whether it is required to be weakly interacting as a picture of DM. 
Commutatively deformed classical gravitation might also have something interesting to say about information flow and entropy 
at black hole event horizons, where nonlocality might
be physically significant.\cite{Solodukhin}  Additionally, one could apply the deformations to other classical gravitational 
actions such as the the first order (Palatini) formulation of the Einstein-Hilbert action, the Holst action, or the Einstein-Cartan 
theory where torsion plays a role. If commutatively deformed general relativity does describe classical spacetime, 
then it would replace the undeformed version as the suitable classical limit of quantum theories of gravity. 
This first classical step towards deformed gravitation is UV incomplete, and its quantum version should be explored.  \\

Finally, from an experimental point of view, the theory is extremely difficult to test, precisely because its energy scale is nearly Planckian.   
For example, Big Bang nucleosynthesis (BBN) and the observed primordial abundance of elements $A=2-7$ are highly sensitive to the value of the time 
derivative of the cosmological scale factor $a$ during that epoch through the Friedmann equations for the Friedmann-Lema\^itre-Robertson-Walker 
cosmology.\cite{Kusakabe} The twist will perturb those equations, and the observed isotopic abundances might be used
to provide further constraints on $M_{\phi}$ and $M_{\lambda}$. 
Regrettably, a simple order of magnitude estimate reveals that if the nonlocality scale 
$\xi_c$ lies in the range $10^2\, -\, 10^5\, L_P$, then the relative perturbations to the Friedmann equations at BBN are only $10^{-88}\,-\,10^{-79}$, 
producing changes in abundances well within the observational error bars. 
The twist is simply too small to measurably affect BBN. \\

An alternative route might be to rule out experimentally the possibility that twist producing matter is comprised of SM baryons.
Using the estimate (\ref{XDPhi_xi_est}), one finds if \emph{all} (and equivalently by the symmetry considerations of section 5, \emph{any}) 
SM baryons generate a twist through $\mathscr{Q}_{l}= B_l$,
then baryons having rest masses $M$ exceeding a non-locality mass $M_{\mathrm{NL}}$ 
will find it problematic to act according to a local field theory that respects microcausality. 
This is because for $M\gtrsim M_{\mathrm{NL}}$, the particle's own twist produced nonlocality scale $\xi_c (M)$ becomes
a significant fraction of its Compton wavelength $\Lambda_{C}(M)$.
Specifically, one obtains
\begin{equation}
M_{\mathrm{NL}} \simeq \left(56.5\;\mathrm{TeV}\right) \left(\xi_c(M)/\Lambda_C(M)\right)^{1/4}\,\left(\xi_c(\mathrm{proton})/L_P \right)^{-1/4}.
\end{equation}
For instance, $\xi_c(\mathrm{proton})/L_P  \simeq 10^2$ and $\xi_c(M)/\Lambda_C(M)\simeq 0.1$ imply $M_{\mathrm{NL}}\simeq 10.0$ TeV, and 
$\xi_c(\mathrm{proton})/L_P \simeq 10^5$ together with $\xi_c(M)/\Lambda_C(M)\simeq 0.1$ yield $M_{\mathrm{NL}}\simeq 1.8$ TeV.
These energies still lie mostly beyond present day accelerator laboratory capabilities; but as baryons of higher rest mass are 
studied and found to continue to behave as law abiding citizens of standard model local quantum field theory, 
then the bounds excluding SM baryonic matter as twist producing particles become tighter. 
Of course such a continuation of baryonic microcausality into the TeV range by itself would not
constitute positive experimental evidence for commutatively deformed general relativity.
The baryons' rest masses $M$ generally (but not
monotonically) increase with their total angular momentum $J$, so one expects eventually to find baryons with $M(J)\gtrsim M_{NL}$, which will violate
microcausality if the baryons are twist producing.
If we anticipate baryonic microcausality to hold through rest masses of a few tens of TeV, that would only leave a right-handed neutrino
or a non-standard GM matter sector as twist generating possibilities. \\

In this article we have studied commutatively deformed diffeomorphisms (diffs)
of curved classical spacetime.
This was motivated by a search for a physical origin for classical nonlocality from which microcausality may 
emerge by the Lieb-Robinson route, despite the fact that many models of background free quantum gravity are acausal even in some 
classical limit.\cite{PdV} The use of coordinate-free Hopf algebra methods maintains the essential background independence of general relativity, 
but at the same time the deformed infinitesimal diffs obey a Lie algebra distinct from the undeformed ones, 
so the theories possess different symmetries. The use of quasi-triangular Hopf algebras ensures that gauge flow in canonical
gravity in the presence of an external time remains anomaly free. 
The twist producing matter could be as familiar as all the SM baryons (which would be acausal for rest masses above $\sim 1-10$ TeV) or a more exotic sterile neutrino.
However, the nonlocally acting deformed diffs may also arise naturally 
from a new sector of matter fields, and there is a range
of their masses and coupling constants that imply a nonlocality length $\xi_c \sim 10^2 - 10^5 $ times the Planck length $L_P$, 
which the Lieb-Robinson mechanism may use to generate microcausality on longer scales.
The commutatively deformed diffs preserve the theoretical architecture of the standard model. 
They also engender presently immeasurable  perturbations of  Big Bang nucleosynthesis, solar system orbits, 
atomic and nuclear spectra, and particle physics branching ratios due to the near Planckian scale of the 
nonlocality. 
In several cases the new sector of matter fields generating the deformed diffs provide viable dark matter candidates. 
If the twist generating matter acts as dark matter, then it will not measurably affect classical electromagnetic radiation propagating 
over cosmological distances. This approach to dark matter makes no appeal to grand unified theories, 
extra dimensions, supersymmetry, strings, mirror worlds, or modifications of Newtonian gravity. \\

\section{Acknowledgements}

The author wishes to thank  S.T. Lu for insightful discussions, as well as Sun Wu Kong and Zhu Ba Jie 
for their creative and imaginative inspiration in Shanghai where the most innovative aspects of this work took place.\\

 \end{document}